# МР-томография в сверхвысоком поле: новые задачи и новые возможности

М.А. Зубков[а], А.Е. Андрейченко[а], Е.И. Кретов[а], Г.А. Соломаха[а], И.В. Мельчакова[а], В.А. Фокин[б], К.Р. Симовский[а,в], П.А. Белов[а], А.П. Слобожанюк[а]

[а]Университет ИТМО, Кронверкский пр. д.49, Санкт-Петербург, 197101, Российская Федерация

[б]ФГБУ «НМИЦ им. В. А. Алмазова» Минздрава России, ул. Аккуратова, д. 2, Санкт-Петербург, 197341, Российская Федерация

[в]Aalto University, School of Electrical Engineering, PO Box 11000, Aalto, FI-00076, Finland

**Аннотация**

Центральная тенденция развития современной магнитно-резонансной (МР) томографии – повышение величины индукции статического магнитного поля, в котором производится регистрация МР-изображений. Использование в МР-томографии сверхвысоких полей (7 Тл и более) сопряжено с множественными эффектами, как повышающими или снижающими качество получаемых изображений, так и не наблюдаемыми в более слабых полях. В данном обзоре представлены наиболее значительные следствия использования сверхвысокого поля в МР-томографии, включающие в себя новые задачи и их решения, а также новые методы и особенности МР-сканирования, нехарактерные для МРТ в низких и высоких полях (менее 7 Тл).

## 1. Введение

Магнитно-резонансная томография (МРТ) – метод исследования внутренней структуры объектов путём наблюдения сигнала ядерного магнитного резонанса (ЯМР) от этих объектов в магнитном поле специально сформированной неоднородности. История МРТ насчитывает уже более половины столетия [1] в России и чуть меньше – за рубежом [2]. Новые теоретические разработки и технологии постепенно меняют облик МРТ, как области знания, генерируя новые принципы получения магнитно-резонансных (МР) изображений (называемых в России томограммами), новые области применения МРТ и схемы обработки полученных изображений. Центральной тенденцией развития МРТ всегда являлось стремление перенести клинические МР-исследования сначала из слабого или, как принято говорить у специалистов по МРТ, *низкого* магнитного поля (до 1 Тл) в так называемое *высокое* поле (от 1 до 3 Тл), а в последнее время – в *сверхвысокое* (более 3 Тл).

Спектр клинических МР-исследований был значительно расширен благодаря переходу от технологий низкого магнитного поля к технологиям высокого поля. Использование высокого поля позволило улучшить временное разрешение и уменьшить длительность исследования, что особо важно для МР-визуализации движущихся органов, таких как сердце, органы брюшной полости (печень, кишечник). Оказалось возможным выполнять исследования печени и других паренхиматозных органов на одной задержке дыхания с высоким пространственным разрешением и контрастом мягких тканей, что особенно важно в диагностике онкологических заболеваний [3]. Благодаря улучшению временного разрешения появилась возможность выполнять такие требовательные к скорости получения изображений исследования, как морфологическую и функциональную диагностику сердца. Последняя производится путём получения изображений в различные фазы сердечного цикла [4]. Функция сердца оценивается численно методом определения фракции сердечного выброса: объема и процента крови, покидающей сердце за единицу времени или за одно сердечное сокращение. В высокопольной МРТ также появилась уникальная возможность использовать парамагнитные контрастные вещества для оценки кровоснабжения сердечной мышцы – миокарда при помощи динамического исследования – МР-перфузии [5]. МР-перфузия активно применяется при оценке кровоснабжения не только сердца, но и головного мозга. Высокопольная МРТ (а в особенности, МРТ в поле 3 Тесла) позволяет исследовать перфузию и без использования контрастных веществ, помечая радиочастотной меткой притекающую по сосудам шеи кровь и затем вычисляя скорость мозгового кровотока в тканях головного мозга [6]. Такая методика артериальных спиновых меток (arterial spin labeling или ASL) активно входит в диагностику различных неврологических и психиатрических заболеваний за счет отсутствия лучевой нагрузки и неинвазивности МР-исследования. Ограничения на частоту проведения таких МР-исследований значительно ниже, чем на аналогичные, использующие ионизирующее излучение.

Только с появлением высокопольной томографии стали доступны новые методики в изучении физиологических процессов в головном мозге – функциональная МРТ (фМРТ), МР-морфометрия и визуализация нервных проводящих путей (МР-трактография). Функциональная МРТ позволяет за счет эффекта зависимости контраста от оксигенации крови (blood oxygenation level dependent или BOLD-контраста) выявлять зоны функциональной активности кровотока в коре головного мозга, как при подаче внешних стимулов, так и в покое для установления связей между различными участками головного мозга [7]. МР-морфометрия – метод, использующий полуавтоматическую сегментацию и вычисление объема различных структур головного мозга [8] и позволяющий проводить дифференциальную диагностику при различных неврологических заболеваниях, например, при различных видах деменций (болезнь Альцгеймера, сосудистые деменции и др.). МР-трактография позволяет на основе данных так называемого диффузионно-тензорного картирования (diffusion tensor imaging или DTI) определить направление преимущественного движения молекул воды, происходящего вдоль миелиновых оболочек проводящих путей [9], и осуществить визуализацию нервных волокон головного мозга и других органов, что находит своё применение как в нейроонкологии, так и в фундаментальных нейрофизиологических исследованиях.

Отдельно стоит отметить развитие МР-спектроскопии на ядрах водорода, фосфора и других элементов, позволяющей проводить неинвазивную диагностику на биохимическом уровне, в частности, благодаря оценке содержания холина – как маркера сохранности клеточных мембран, лактата – как маркера анаэробного гликолиза при недостаточном поступлении кислорода, N-ацетил аспартата – как маркера активности нейронов, а также ряда других метаболитов [10]. Фосфорная высокопольная МР-спектроскопия позволяет оценить энергетические процессы, протекающие в печени и мышцах, путём оценки содержания аденозинтрифосфата (АТФ) и аденозиндифосфата (АДФ) [11,12].

Развитие высокопольной томографии также позволило улучшить неинвазивную визуализацию сосудов благодаря использованию фазово-контрастных методик, в которых движущаяся кровь обладает собственным естественным контрастом. При таких исследованиях в высоких магнитных полях становится возможным выполнять 4D–визуализацию кровотока (при изучении ламинарного и турбулентного кровотоков), для планирования сердечно-сосудистых операций [13]. В изучении опорно-двигательного аппарата при использовании высокопольной МРТ стала возможной визуализация строения хряща, а также обнаружение дегенеративных изменений до появления видимых нарушений [14].

Таким образом, внедрение высокопольной МРТ в клиническую практику позволило не только улучшить анатомическую визуализацию различных областей тела, но и оценить физиологические и даже биохимические процессы, протекающие в организме. При этом вреда для организма от использования магнитного поля на уровне 3 Тл выявлено не было. Сверхвысокопольная МРТ открывает еще более широкие горизонты. Несмотря на то, что целесообразность дальнейшего увеличения индукции магнитного поля в МРТ иногда оспаривается [15], переход как клинической, так и доклинической МРТ в область сверхвысокого поля представляется нам вполне оправданным и даже неизбежным ввиду целого ряда преимуществ по сравнению с высокопольной МРТ. Некоторые преимущества сверхвысокопольных технологий можно объяснить непосредственно, исходя из принципа получения МР-изображений, другие достоинства следуют из зависимости свойств исследуемого объекта от величины магнитного поля. Разумеется, у сверхвысокопольных технологий имеются и недостатки, на которых и основана их критика [15]. Преимущества и недостатки сверхвысокопольных технологий будут описаны в следующем разделе статьи. Так как данная статья предназначена для широкого круга физиков, в том числе незнакомого с МРТ, обсуждение особенностей сверхвысокопольной МРТ возможно лишь после краткого изложения основных принципов МРТ как одного из методов экспериментальной физики.

## 2. Особенности сверхвысокопольной МР-томографии

Принципиально МР-томография является частным случаем ЯМР-спектроскопии, в основе которой лежит явление ядерного магнитного резонанса (ЯМР). Как известно, у атомных ядер с ненулевым спином в статическом магнитном поле возникает зеемановское расщепление энергетических уровней. Например, для спина протона (водород входит в состав молекулы воды именно в виде двух протонов), равного 1/2, возникает два уровня. Разность энергий верхнего и нижнего уровней пропорциональна величине статического поля $B_0$. Если к объекту, содержащему воду или другое вещество, у которого атомные ядра имеют ненулевой спин, приложить радиочастотное (РЧ) магнитное поле, то возникают неравновесные квантовые переходы, которые имеют резонансный характер, если частота поля равна частоте прецессии ядер (ларморовской частоте). Такое РЧ поле на ларморовской частоте создается передающими антеннами, которые в исследованиях по МРТ называются катушками, даже если антенная система представляет собой решетку дипольных антенн. Этот термин нуждается в пояснении. Если в дальней зоне излучения антенной системы напряженности магнитного и электрического полей равны друг другу с точностью до волнового импеданса пространства, то в ближней зоне магнитное поле антенны может доминировать над электрическим. Так как расстояние от объекта до антенной системы в МР-томографии меньше длины волны, а электрическое РЧ поле антенны представляет собой паразитный эффект, ибо не создает полезного сигнала, а лишь ведет к индукционному нагреву биологических тканей, то антенные системы в МРТ нацелены на эффективное создание именно ближнего магнитного поля. Поэтому термин РЧ катушка является оправданным в МРТ.

Итак, первичное магнитное поле ларморовской частоты, созданное РЧ-катушкой, приводит к возбуждению верхнего зеемановского состояния атомных ядер в объекте, которые затем возвращаются к нижнему состоянию, причем в процессе восстановления равновесного состояния испускается вторичное излучение с той же несущей частотой. Взаимодействие этого излучения либо с отдельными приемными РЧ-катушками, либо с теми же приемо-передающими катушками (что возможно ввиду задержки по времени относительно первичного РЧ-импульса) является основой для возникновения регистрируемого сигнала ЯМР [16]. Особенностью МРТ (как подвида ЯМР-спектроскопии) является намеренное кратковременное создание статического магнитного поля различной величины в разных местах исследуемого объекта (обычно, с линейной пространственной неоднородностью). Для этого на располагающиеся по периметру МРТ-туннеля специальные катушки постоянного тока – *градиентные катушки*, подаются импульсы постоянного тока, так называемые импульсы градиентов. Вследствие наличия такой неоднородности статического поля сигнал ЯМР, возбуждаемый в приёмной РЧ-катушке излучением от разных мест объекта, отличается не только по амплитуде (ввиду различной концентрации атомных ядер данного вещества в разных местах объекта), но и по частоте. В итоге можно соотнести амплитуду сигнала ЯМР данной частоты с координатами соответствующей области внутри объекта, что и дает картину пространственного распределения по объекту массовой плотности исследуемого вещества, например, воды, жиров или фосфора [7,17]. Дополнительный анализ фазовых соотношений между несколькими сигналами ЯМР, регистрируемыми после подачи определенной последовательности РЧ-импульсов и импульсов градиентов [7,17], как раз и позволяет получить в МРТ детальное изображение различных сечений объекта, но описание таких, пусть и важных, деталей метода не является целью данного обзора.

Для дальнейшего рассмотрения важно отметить, что вследствие квантовой природы эффекта ЯМР обладает такими характеристиками, как узкополосность и пропорциональность частоты наблюдаемого сигнала величине индукции статического магнитного поля ($B_0$). Впрочем, несмотря на квантовую природу, явление ЯМР в большинстве случаев может быть описано классически – как индуцированную РЧ магнитным полем прецессию макроскопической ядерной намагниченности вокруг статического магнитного поля. Согласно этой модели, воздействие РЧ поля приводит к отклонению вектора макроскопической намагниченности от равновесного направления – направления статического магнитного поля. Разумеется, из сказанного понятно, что для наблюдения ЯМР поле ларморовской частоты должно иметь направление магнитного вектора, отличное от статического. Для максимально эффективного взаимодействия с прецессирующим вектором макроскопической ядерной намагниченности РЧ поле (в МРТ оно обозначается $B_1^+$) должно иметь круговую поляризацию в плоскости, ортогональной вектору **B**$_0$. Чаще всего ЯМР в этом случае описывают уравнениями Блоха – уравнениями вынужденной прецессии магнитного момента в стационарной или во вращающейся системе координат. Эти уравнения представляют собой полуклассическую модель эффекта. В уравнения Блоха, в зависимости от проводимого эксперимента, могут добавляться или не добавляться релаксационные, градиентные или диффузионные члены [7–9,17].

С другой стороны, в максимально достижимых на данный момент магнитных полях резонансная частота большинства изотопов с ненулевым магнитным моментом попадает в радиочастотный диапазон, а значит МРТ может быть рассмотрено, как подраздел радиофизики. С радиотехнической точки зрения ЯМР-спектроскопия и МР-томография могут быть разделены на две последовательные задачи. Первая это – создание однородного радиочастотного магнитного поля в заданной области, вторая – приём, усиление и регистрация слабого узкополосного сигнала, находящегося в ближнем поле одной или нескольких антенн; при этом приёмные антенны могут, как совпадать с передающими, так и отличаться от них. Величина индукции магнитного поля не влияет непосредственно на радиотехническую часть оборудования для наблюдения ЯМР, однако то, что она определяет рабочую частоту всей РЧ части спектрометра или томографа приводит к существенной зависимости параметров приёма и передачи РЧ энергии от величины статического магнитного поля, в котором производится наблюдение явления ЯМР. В частности, классическим следствием повышения величины постоянного магнитного поля является квадратичное повышение амплитуды сигнала ЯМР, а следовательно – и чувствительности МР-томографии [17]. Поэтому сверхвысокопольная МРТ позволяет получать изображения с таким качеством, которое недостижимо для высокопольных технологий. Тем самым открываются новые возможности экспериментального исследования головного мозга, внутренних органов, суставов, нервной системы и т.д.

Однако более детальный анализ МР-томографии с точки зрения радиотехники показывает, что повышение постоянного магнитного поля выше 3 Тл ведет и к негативным последствиям: росту нагрева тела исследуемого пациента или образца биологических тканей при повышении $B_0$ [18] и усилению пространственной неоднородности РЧ магнитного поля [19]. Оба эффекта связаны с ростом частоты наблюдения сигнала ЯМР, пропорциональной величине $B_0$. Усиление пространственной неоднородности естественным образом следует из сопутствующего сокращения рабочей длины волны в полях выше 3 Тл. Действительно, хотя размеры исследуемой области тела или образца тканей как правило меньше длины волны электромагнитного поля в свободном пространстве, эти размеры (в высокопольных и особенно в сверхвысокопольных технологиях МРТ) часто превышают длину волны внутри тела. Дело в том, что биологическая ткань в значительной степени состоит из воды – материала, который на радиочастотах имеет высокий показатель преломления, то есть волна, преломленная внутрь ткани, существенно укорачивается. В итоге, внутри тела возникает интерференция волн, испытывающих внутреннее отражение, иными словами образуется стоячая волна. Как известно, в стоячей волне электромагнитное поле существенно неоднородно ввиду чередования узлов и пучностей. С ростом $B_0$, то есть с укорочением длины волны, масштаб характерной неоднородности поля очевидным образом, уменьшается. Если масштаб неоднородности РЧ поля оказывается сравним с размером визуализируемых структур в исследуемом образце, изображения этих структур будут искажены. Таким образом, две центральные радиотехнические проблемы сверхвысокопольной МРТ – это неоднородность первичного (и вторичного!) РЧ поля, возникающие вследствие интерференции электромагнитного излучения, внутри объекта исследования, а также повышенное поглощение РЧ-энергии объектом, приводящее к его нагреву [15,20,21].

С точки зрения релаксационной теории в простейшем ЯМР эксперименте происходит нарушение равновесного распределения частиц по энергетическим уровням с последующим (как по окончании РЧ импульса, так и частично во время его действия) восстановлением равновесного состояния. В данном случае величина статического магнитного поля непосредственно влияет на скорость обмена энергией как между спинами, так и на скорость обмена энергией спинов с окружающей средой. В томографии эти скорости принято характеризовать константами $T_1$, $T_2$ (временами спин-решёточной и спин-спиновой релаксации соответственно) и $T_2^*$, включающей также влияние окружающей (в микроскопическом масштабе) среды на поперечную ядерную намагниченность. Зависимость $T_1$, $T_2$ и $T_2^*$ от величины $B_0$ неоднозначна и зависит от рассматриваемого вещества. В контексте клинической томографии наличие данной зависимости приводит в первую очередь к изменению разности сигналов (контраста) от различных тканей, а также к изменению длительности экспериментов вследствие изменения времени возвращения макроскопической намагниченности к состоянию равновесия [22]. Разница в величине $T_1$ ведет к новым диагностическим возможностям сверхвысокопольной томографии, которые для низкопольных и даже высокопольных технологий недоступны [23].

Однако, сложности, связанные с неоднородностью РЧ поля требуют разработки новых методов сканирования, позволяющих получить корректное отображение распределения времен релаксации $T_1$ в исследуемой анатомической области без превышения допустимых норм РЧ облучения пациента [24,25]. Увеличение величины $T_1$ и повышенный риск нагрева пациента мешают получить классическими методами корректную $T_2$-взвешенность, поэтому для сверхвысокопольного сканирования также разрабатываются новые методы получения $T_2$-взвешенных изображений [26–28]. Изменения в контрасте по $T_2^*$ приводит к значительному изменению целого направления МРТ – функциональной магнитно-резонансной томографии (фМРТ), которой будет посвящен отдельный раздел данного обзора.

Даже такой беглый обзор особенностей проведения МР-исследований в сверхвысоком поле показывает, что увеличение $B_0$ является значимым, хотя и непростым, шагом для современной МРТ. Этот шаг приводит к появлению новых возможностей и задач МР-томографии, к смене технических подходов к МР-сканированию, и в некоторых случаях – к сдвигу парадигмы проведения некоторых практических МР-исследований. Последующие части данного обзора посвящены наиболее существенным и значимым техническим и методологическим решениям и нововведениям в МРТ, вызванным появлением сверхвысокопольных МР-томографов для сканирования человека.

## 3. Объёмные и многоканальные системы приёма и передачи

Традиционными объёмными РЧ-катушки для возбуждения ядерной спиновой системы в МРТ являются кольцевые резонаторы, коаксиальные с МРТ-туннелем. В низкопольных (менее 1.5 Тл) и высокопольных (от 1.5 до 3 Тл) МРТ-технологиях наиболее известны резонаторы типа «птичья клетка» [29] и ТЕМ-резонатор [30]. В этих объемных катушках ток течет по проводникам вдоль оси туннеля, в то время как в азимутальных направлениях образуется стоячая волна. При ограничении геометрических размеров катушки размерами туннеля томографа для совпадения длины стоячей волны с длиной волны сигнала ЯМР необходимо значительное замедление первой, что достигается с использованием либо конденсаторов большой емкости [29] либо емкостей, возникающих между концентрическими слоями проводов [30]. Оба вышеупомянутых резонатора позволяют создать максимально однородное (по крайней мере в отсутствие объекта) первичное РЧ поле внутри катушки. Если исследуемый объект центрируется на оси катушки, то в его центре РЧ поле оказывается достаточно велико и соответствует пучности стоячей волны магнитного поля, которая, как уже отмечалось, возникает за счет внутреннего отражения поля в объекте. Узлы стоячей волны находятся на расстоянии примерно равном четверти длины волны от пучностей. На частотах менее 130 МГц (которые соответствуют постоянным магнитным полям 3 Тл и ниже), длина волны в тканях при типичной величине их диэлектрической проницаемости $\varepsilon_r \approx 50\text{-}60$ [31] относительно велика в сравнении с размером, например, мозга человека. В этом случае узлы стоячей волны внутри сканируемого объёма не возникают. При уровне постоянного поля 7 Тесла (что соответствует рабочей частоте примерно 300 МГц) длина волны в мозге человека составляет величину порядка 13 см, т.е. сравнима с размерами области сканирования. Как уже отмечалось выше, это приводит к возникновению артефактов на МР-изображении. Поэтому традиционные РЧ-катушки не считаются перспективными для сверхвысокопольной МРТ. Здесь необходимо отметить, что пока еще не до конца ясно, насколько существенно влияние диэлектрической проницаемости тканей на неоднородность поля $B_1^+$ [32]. Дело в том, что существует еще один фактор, влияющий на однородность поля $B_1^+$ – это большая электрическая проводимость биологических тканей, которая с трудом поддается измерению и может зависеть от ряда трудно учитываемых факторов. Из-за высокой электропроводности влияние диэлектрических эффектов снижается, и на практике распределения поля $B_1^+$ получаются более однородными [33]. Тем не менее, проблема выравнивания распределения магнитного поля внутри объекта является ключевой для сверхвысокопольной МРТ.

Создание более однородного радиочастотного поля в системе сверхвысокопольной МРТ (в сравнении с полем, создаваемым традиционными объёмными катушками) возможно если катушка представляет собой такой набор антенных (проводников) элементов, в котором фаза и амплитуда сигнала для каждого проводника контролируется индивидуально и оптимизируется с учетом влияния объекта на РЧ магнитное поле. Процесс оптимизации фаз и амплитуд сигналов, поданных на отдельные провода такой катушки, называется РЧ-шиммированием [34,35]. Этот процесс осуществляется путём регистрации пространственного распределения поля, создаваемого отдельными антенными элементами, и последующего вычисления (с использованием полученных распределений) требуемого набора фаз и амплитуд тока для создания максимальной однородности поля в выбранной области. На практике нечто подобное осуществляется и в объёмных катушках типа «птичья клетка» [29] и ТЕМ-катушка [30]. Если любую из них рассматривать как резонатор, то в ней существуют две линейно поляризованные собственные моды. Две данные моды можно возбудить с фазовым сдвигом 90°, что будет соответствовать полю круговой поляризации внутри катушки. Однако, при наличии двух независимых каналов питания, эти две моды можно возбудить и с любой другой разностью фаз. Подбор разности фаз отличной от 90° может помочь скомпенсировать вызываемое исследуемым объектом отклонение поля от круговой поляризации, в результате чего спиновая система самого объекта будет возбуждаться полем максимально близким к круговой поляризации, что ведет к улучшению качества получаемого изображения. Такая оптимизация фазы является простейшим типом РЧ-шиммирования, который получил широкое распространение в современных томографах с величиной индукции магнитного поля 3 Тл, и в частности при исследованиях органов грудной и брюшной полости человека [36]. Однако, в сверхвысоких полях для исследования мозга и органов, находящихся на большой глубине в теле человека, РЧ-шиммирование с использованием двух сигнальных каналов является недостаточным. Для снижения неоднородности РЧ поля в этом случае необходимо увеличивать число каналов. Процесс возбуждения спиновой системы множеством независимых сигнальных каналов называется параллельной передачей [37]. Суммарное распределение РЧ поля многоканальной катушки, состоящей из независимо питаемых антенных элементов, может быть сделано существенно более однородным, чем поле двухканальной катушки аналогичного размера.

При создании таких многоканальных катушек – по сути дела антенных решеток, у которых кроме распределенной фазы распределена также и амплитуда – особо актуальной становится проблема развязки отдельных антенных элементов. Взаимные влияния элементов в таких антенных решетках чрезвычайно сильны, потому что расстояния между соседними проводами оказывается меньше длины волны в 10 и более раз. В отличие от обычных антенных решеток, взаимодействие элементов такой катушки приводит не столько к расстройке решетки и искажению ее диаграммы направленности (что мало существенно для МРТ), сколько к практической невозможности шиммирования. Дело в том, что требуется оптимизировать фазы и амплитуды токов в проводниках, которые питаются от реальных источников сигнала со специально подобранной амплитудой и фазой. Наиболее адекватной моделью питания элемента решетки является генератор напряжения с конечным (известным) выходным импедансом. Слишком сильные взаимные влияния элементов приводят к тому, что соотношения между токами в антенных элементах и напряжениями, приложенными к ним, становятся практически непредсказуемыми, а потому токи в элементах решетки становится невозможно контролировать. Поэтому при применении антенных решеток для РЧ-шиммирования необходимо снижать эти взаимные влияния до приемлемых величин.

На данный момент в качестве антенных элементов катушек, работающих в режиме параллельной передачи, в сверхвысокопольной МРТ используются рамочные антенны [38], резонансные отрезки длинных линий [39] и дипольные антенны [40]. Для задач по исследованию мозга в полях 7 и более Тесла лучше всего зарекомендовали себя катушки, представляющие собой массивы рамочных антенн [41] (Рис. 1, а, б, в), а в задачах по исследованию человеческого тела – комбинации диполей и рамочных антенн [42] (Рис. 1, г, д, е).

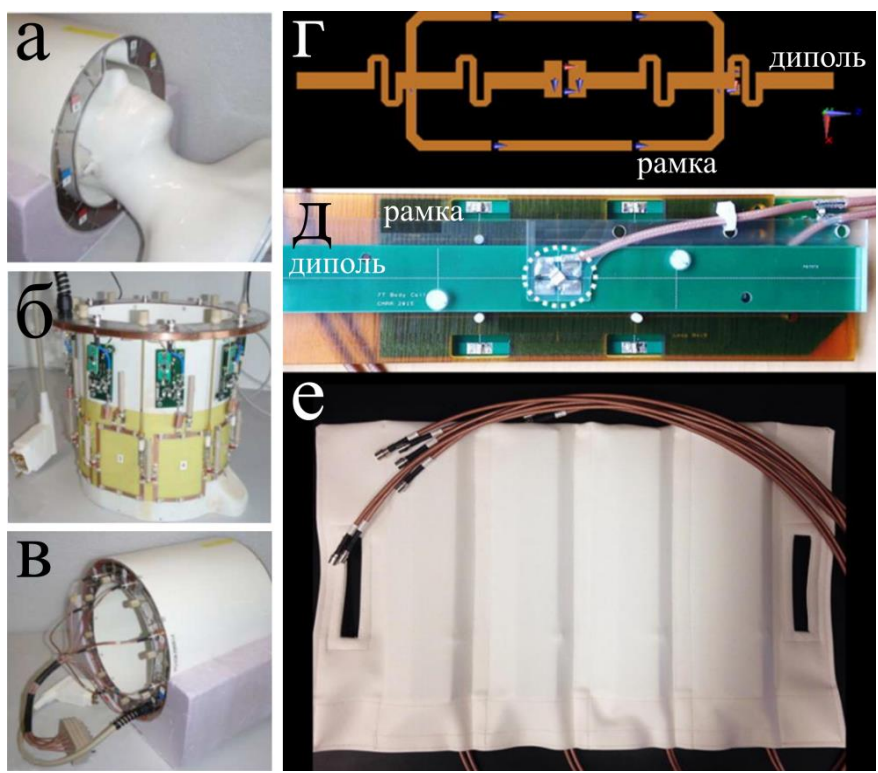

**Рис. 1 (а), (б), (в) – восьмиканальная РЧ-катушка для исследования человеческого мозга. (г, д) – элемент 16-ти канальной катушки на основе комбинаций диполей и рамок для исследования человеческого тела на поле 7 Тл, (е) – верхняя часть 16-ти канальной катушки для тела на основе комбинации диполей и рамок.**

Вопросы развязки подобных антенн для применения в МРТ интенсивно исследуются. Для рамочных антенн главными причинами взаимной связи являются взаимная индуктивность и резистивная связь через объект сканирования. Взаимная индуктивность двух рамок может иметь как положительное значение, так и отрицательное. Это свойство используется для самой распространённой техники развязки рамок – за счёт частичного перекрытия соседних элементов. При правильном подборе области перекрытия взаимные индуктивности перекрывающихся и не перекрывающихся частей двух рамок компенсируют друг друга [43]. Иногда для лучшей развязки в таких катушках используют дополнительные пассивные рамки [44]. Широко известны и другие методы развязки рамок, например, с помощью конденсаторов [45], включаемых в одновременно в плечи рамок, или с помощью трансформаторов [46]. Методы развязки для РЧ-катушек, представляющих собой массивы электрических диполей на данный момент недостаточно хорошо разработаны. Один из возможных подходов к этому вопросу - размещение пассивного диполя между двумя соседними активными [47]. Главным недостатком этого метода является сильное влияние пассивного диполя на распределение поля в теле человека. Существует и другой метод развязки дипольных антенн, находящихся на расстоянии порядка $\lambda/10$ [48]. Развязка в данном случае осуществляется с помощью размещения антенной решетки на поверхности структуры, представляющей собой двумерный аналог фотонного кристалла, причем построенной таким образом, что рабочая частота катушки попадает в запрещенную зону структуры [49].

При параллельной передаче суперпозиция полей антенных элементов может происходить либо в так называемом статическом режиме, когда соотношение токов в сигнальных каналах антенной решетки остаётся неизменным в ходе излучения РЧ-импульсов, либо в динамическом режиме. В последнем случае сигналы отдельных каналов варьируются от импульса к импульсу. Это позволяет достичь высокой пространственной однородности угла отклонения макроскопической ядерной намагниченности относительно направления **B**$_0$. Однородность достигается благодаря использованию многократного возбуждения короткими импульсами при малых углах отклонения (то есть при низких амплитудах РЧ поля) с различными вкладами каналов в ходе излучения каждого импульса [50]. Использование статического режима параллельной передачи применяется, в основном, для сверхвысокопольного сканирования мозга [41,51,52], сердца [53–55] и других органов [56–60]. В динамическом режиме [61] каждый радиочастотный импульс низкой амплитуды служит для получения информации от некоторой части пространства. Имеется в виду как физическое пространство, причём выделение пространственной области достигается путём перераспределения взвешенности каналов, так и фазовое пространство. В последнем случае для выделения области в пространстве волновых векторов используются импульсы градиентов магнитного поля, включаемые в промежутках между РЧ-импульсами. Такой подход предоставляет гораздо больше степеней свободы для контроля однородности РЧ-поля. В динамическом режиме использование даже одного дополнительного возбуждения позволяет достичь большей суммарной однородности финального изображения, чем в случае статической параллельной передачи [62–64]. Следует отметить, что целью процесса оптимизации может быть не только максимальная однородность угла отклонения намагниченности в заданной области пространства, но и уменьшение риска перегрева пациента путём оптимизации распределения электрического поля.

Для приёма РЧ-сигналов в клинических МРТ-системах также обычно применяют многоканальные массивы приёмных катушек. Стандартное количество приёмных каналов в них составляет 16 или 32 (Рис. 2, а), хотя в исследовательских целях иногда создаются массивы с гораздо большим числом каналов, например 96 [65]. Мотивацией к увеличению числа каналов служит преобладание в высоких и сверхвысоких полях шума, генерируемого телом человека, над шумом приёмного канала. Вследствие этого в режиме приёма фазированная решётка может обеспечить лучшее отношение сигнал/шум по сравнению с объёмными катушками (Рис. 2, б, в, г). Это достигается за счёт того, что каждый элемент приёмной катушки практически принимает шумовой сигнал только из области непосредственно примыкающей к этому элементу катушки. Для получения максимального отношения сигнал/шум, как и в случае с передающими массивами, необходимо чтобы отдельные приемные каналы были развязаны. Для приёмных каналов эта задача значительно проще, потому что взаимная связь может быть уменьшена за счёт использования специально рассогласованных предусилителей [43]. Чтобы не вносить искажений в сигнальное поле $B_1^+$, а также не повредить электронику усилителей, во время передачи приёмные каналы (если антенная система является приемо-передающей) или приемная катушка (если она существует отдельно от передающей) отключаются с помощью системы p-i-n диодов.

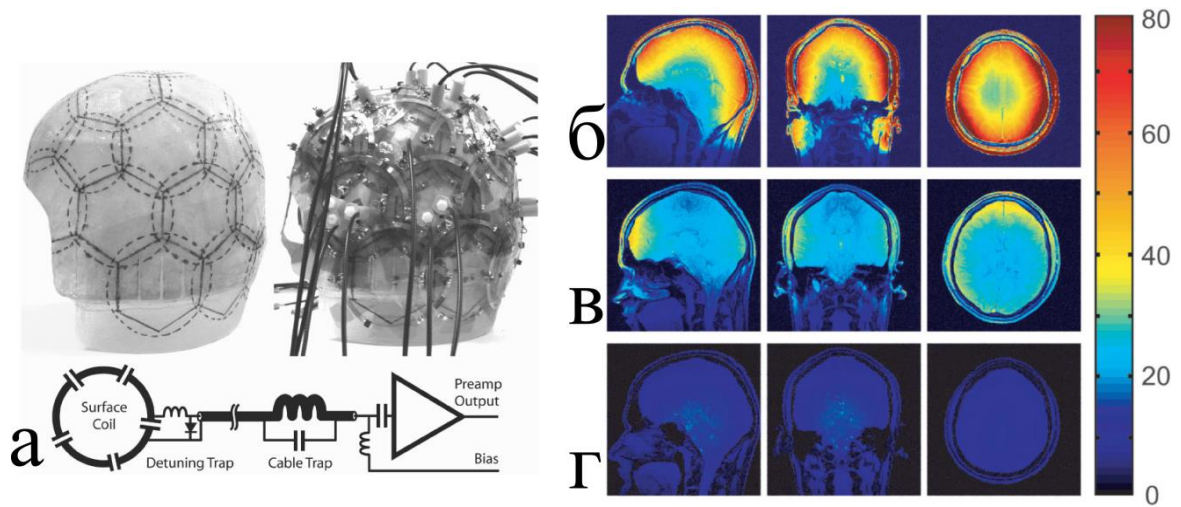

Рис. 2 (а) – 32-х канальная катушка для исследования мозга при уровне поля 3 Тесла и электрическая схема одного приёмного элемента. Распределение отношения сигнал/шум в голове человека для различных приёмных катушек: (б) – 32-х канальная катушка, (в) – коммерческая 8-ми канальная катушка, (в) – коммерческая объёмная катушка для головы.

## 4. Метаматериалы и диэлектрические структуры в МРТ

Одним из наиболее передовых направлений в развитии технологии сверхвысокопольной МРТ является использование метаматериалов, диэлектрических подкладок и резонаторов. Метаматериалы представляют собой композитные структуры, как правило, состоящие из геометрической комбинации проводников и диэлектриков [66–70]. Характерная особенность метаматериалов в том, что составляющие их элементы, а также расстояния между ними много меньше длины волны электромагнитного излучения, с которым эти элементы взаимодействуют. Как следствие, с помощью метаматериалов можно создавать среды как с высокими величинами эффективной диэлектрической ($\varepsilon$) и магнитной ($\mu$) проницаемостей, так и с отрицательными значениями этих материальных параметров [71]. Хотя подобные свойства метаматериалов имеют место лишь в сравнительно узком диапазоне частот, это не является проблемой в МРТ, где регистрируемые сигналы имеют относительную полосу частот порядка одного процента и менее. Так как искусственный магнетизм (на ларморовской частоте) не требует использования магнитных материалов, то метаматериал с высоким или даже отрицательным значением μ на ларморовской частоте может быть изготовлен из немагнитных компонентов, например меди и пластмассы, что позволяет свести практически к нулю паразитное взаимодействие со статическим магнитным полем $B_0$.

Конструкции на основе метаматериалов можно использовать совместно со стандартными катушками аппаратов МРТ для перераспределения компонент ближних РЧ-полей антенных элементов наиболее эффективным (с точки зрения величины отношения сигнал-шум в области исследования) образом, что ранее было продемонстрированно для МРТ систем 1.5 Тл [72–74]. В силу конструкционных особенностей, свободное пространство внутри аппарата МРТ часто весьма ограничено, поэтому устройства из метаматериалов стараются делать небольших размеров и как можно более тонкими, заменяя объемные метаматериалы на метаповерхности [75,76]. Метаповерхность, как следует из названия - не что иное, как тонкий слой метаматериала. Распространенный тип метаповерхности, используемый как в высокопольной [77–79], так и в сверхвысокопольной [80–84] МРТ – плоская решетка параллельно расположенных проводов [85]. Подобный массив можно рассматривать в качестве слоя анизотропной среды с большим значением эффективной диэлектрической проницаемости для электромагнитных волн, поляризованных в плоскости проводников и распространяющихся в поперечном к ним направлении. Ближние поля массива перераспределяются таким образом, что в центральной области структуры уменьшается электрическое поле, и в тоже время усиливается магнитное. Применительно к МРТ это означает увеличение отношения сигнал/шум в определенной области пространства. Проводники при этом должны быть настроены на полуволновой резонанс. Чтобы добиться компактности метаповерхности, предлагается комбинация из цельных проводов, чередующихся с проводами, имеющими периодические диэлектрические вставки [80]. Если расположить такую метаповерхность внутри стандартной катушки, применяемой для обследований головного мозга, она не создаст паразитного электрического поля в своей окрестности, и потому не причинит дискомфорта пациенту, проходящему процедуру сканирования, и в тоже время позволит улучшить характеристики изображения за счет значительного повышения магнитного РЧ поля в нужной области пространства (Рис. 3). Дополнительные преимущества, например динамическая перестройка свойств метаповерхности, может быть обеспечена путем использования концепций нелинейных и перестраиваемых метаматериалов [86–88].

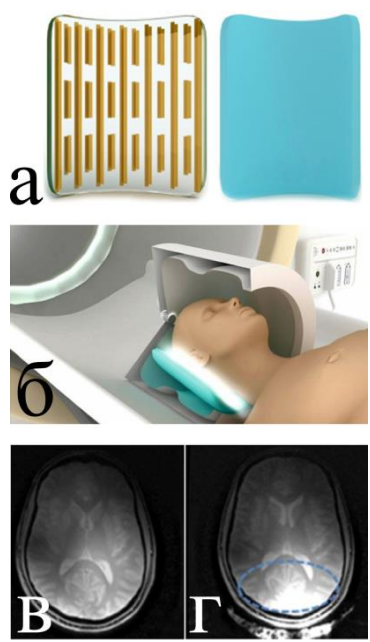

**Рис. 3 Гибкая метаповерхность для перераспределения РЧ-поля в сверхвысокопольной МР-томографии. (а) конструкция метаповерхности, полученной путём комбинации структуры из проводников с диэлектрическими вставками. (б) – расположение устройства внутри стандартной РЧ-катушки МР-томографа при сканировании мозга человека. (в) – изображение мозга человека без использования метаповерхности, (г) – усиление сигнала с помощью метаповерхности (область усиления выделена пунктиром).**

На изображении головного мозга (Рис. 3, в, г) наблюдается вызванное интерференцией РЧ поля неоднородное распределение интенсивности сигнала в объеме среза, выражающееся в виде темных и светлых зон. Метаповерхность из проводов не позволяет устранить этот эффект, так как он вызван внутренним отражением волн от поверхности мозга. Одним из способов борьбы с подобной неоднородностью является применение диэлектрических материалов в МРТ [89], в частности диэлектрических подкладок. В отличие от метаматериалов, подкладки представляют собой более простые устройства – образцы из единого материала со значением относительной диэлектрической проницаемости значительно выше, чем у исследуемого объекта (80>$\varepsilon_r$>1000). На границе объекта с такой подкладкой внутреннего отражения РЧ сигнала не возникает, а потому оказывается возможным переместить узлы стоячей волны из объекта в подкладку. Так, емкость с водой ($\varepsilon_r$≈80), размещенная возле головы пациента позволяет вынести зону деструктивной интерференции из объёма головного мозга [90], и тем самым обеспечить достаточную однородность первичного поля в этой области (Рис. 4, а). Как уже было отмечено, в МРТ подобные процедуры улучшения характеристик однородности поля в зоне обследования называются шиммированием. В данном случае речь идет о пассивном шиммировании.

Кроме воды, хорошо подходящим материалом для пассивного шиммирования является керамика на основе титаната бария *BaTiO₃* и титаната кальция *CaTiO₃* [91]. Диэлектрическая проницаемость этих материалов достигает нескольких сотен и они уже долгое время применяются в немагнитных конденсаторах, используемых в МРТ. В данном случае предлагается формирование больших образцов такого материала в виде подкладок. Присутствие диэлектрической подкладки с $\varepsilon_r$>100 улучшает однородность РЧ-поля в регионе, близком к ее поверхности [92]. В связи с этим появляется возможность снизить амплитуду возбуждающего радиочастотного импульса без существенного ухудшения качества изображения. Это полезно при проведении обследований определенных групп пациентов, на которых РЧ поля могут оказывать вредное воздействие, например людей, использующих кардиостимуляторы. Моделирование сканирования мозга пациента, обладающего подобным устройством, демонстрирует возможность возникновения локальных очагов нагрева тканей рядом с кардиостимулятором опасных для здоровья. Использование подкладки в форме шлема из керамики [93] дает возможность получить достаточную интенсивность РЧ-сигнала в зоне головного мозга, и при этом многократно снизить радиочастотную нагрузку на остальное тело пациента (Рис. 4, б).

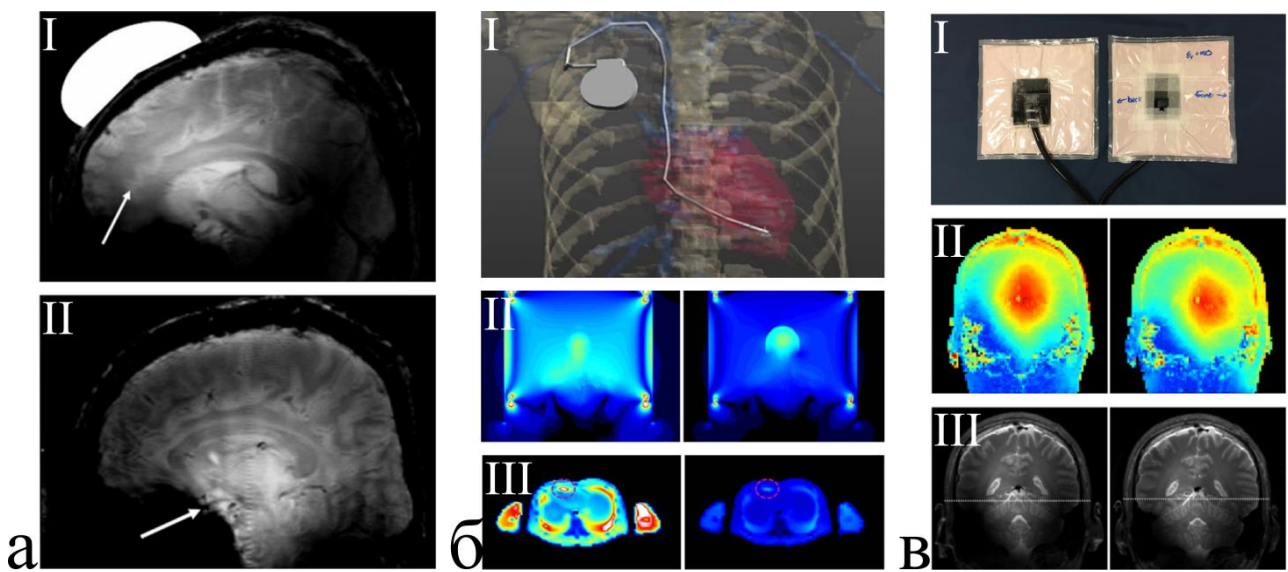

**Рис. 4 Нерезонансные устройства для улучшения качества изображения в сверхвысокопольной МР-томографии. (а) – РЧ-шиммирование с помощью ёмкости с водой: (I) – распределение интенсивности сигнала от мозга человека в присутствии ёмкости (стрелкой отмечена область максимального сигнала), (II) – распределение интенсивности в отсутствие ёмкости со смещенной областью максимальной интенсивности. (б) – перераспределение поля при моделировании сканирования пациента с кардиостимулятором: (I) – модель пациента, (II) – распределение поля в области головы без диэлектрической прокладки (слева) и с прокладкой (справа), (III) – распределение поля в области кардиостимулятора без диэлектрической прокладки (слева) и с**

прокладкой (справа). Кардиостимулятор отмечен пунктиром. (в) – диэлектрические наушники (I) для пассивного РЧ-шиммирования: (II) – распределение поля в модели без наушников (слева) и с наушниками (справа), (III) - распределение интенсивности сигнала в эксперименте без наушников (слева) и с наушниками (справа).

В клинических системах МРТ людям, проходящим обследование, приходится использовать средства индивидуальной защиты органов слуха – наушники. Такая необходимость обусловлена высоким уровнем акустического шума, издаваемого градиентной системой томографа в процессе ее работы. В тоже время наушники необходимы для поддержания связи оператора с пациентом, поэтому они являются обязательным атрибутом практически любой современной клинической системы. Если же использовать в конструкции наушников порошок из керамики с высокой относительной диэлектрической проницаемостью, они могут реализовывать еще и третью функцию – пассивное шиммирование РЧ-поля в области головного мозга [92]. Такой оригинальный способ применения диэлектрических подкладок позволяет улучшить параметры клинической системы, не привнося в ее состав дополнительных конструктивных элементов (Рис. 4, в).

Все представленные на Рис. 4 подкладки представляют собой нерезонансные структуры. Однако, для высокопольных МРТ технологий предлагаются и резонаторы. Использование диэлектрических резонаторов позволяет осуществлять шиммирование в тех случаях, когда простая подкладка не дает результата. Речь идет о шиммировании за счет выбора резонаторных мод. В зависимости от объекта исследования для шиммирования может быть выбрана та или иная мода, позволяющая максимально улучшить однородность поля возбуждения в объекте. Более того, так как в модах диэлектрического резонатора пространственные максимумы электрического и магнитного полей разнесены, то благодаря использованию диэлектрического резонатора можно подавить электрическое поле внутри объекта и усилить магнитное, и тем самым улучшить отношение сигнал/шум. В этом случае резонатор одновременно создает два эффекта - метаповерхности и диэлектрической подкладки. Например, диэлектрический резонатор может быть выполнен в виде трубки (полого цилиндра), причем исследуемый объект при этом располагается внутри полости. У такого резонатора имеются две низкочастотные моды. Одна из них – это так называемая TE-мода, на частоте которой электрическое поле с азимутальной поляризацией практически полностью сосредоточено внутри диэлектрика, а магнитное поле сосредоточено внутри полости, где поляризовано аксиально и имеет максимум на оси трубки. Вторая мода – это так называемая HEM-мода, на частоте которой магнитное поле внутри полости поляризовано перпендикулярно оси трубки и также имеет максимум на этой оси. Частоты этих мод расположены близко, и их резонансные полосы перекрываются, что делает возможным возбуждение обеих мод на частоте перекрытия, причем также достигается круговая поляризация РЧ поля. Помещение биологического объекта в полость трубки лишь незначительно увеличивает неоднородность магнитного поля, особенно если объект достаточно заполняет эту полость [94]. На этом принципе работает одна из предложенных МРТ-катушек для сканирования запястий в сверхвысоком магнитном (Рис. 5, а). Катушка представляет собой трубчатый цилиндр из плексигласа, заполненный дистиллированной водой, причем внутренний диаметр подобран по толщине руки. Получившееся таким образом устройство гораздо проще и дешевле, чем ранее известные многоканальные катушки, используемые в сверхвысокопольной МР томографии. Трубчатый резонатор возбуждается двумя квадратными рамками размером 5 см каждая (две рамки необходимы для возбуждения двух вышеупомянутых мод). Несмотря на небольшой размер рамок, они, тем не менее, эффективно возбуждают резонатор на частотах 298-299 МГц ввиду сильной связи с резонаторными модами. Такая катушка, несмотря на ее простоту, почти не уступает многоканальным катушкам по качеству получаемых изображений запястья [94].

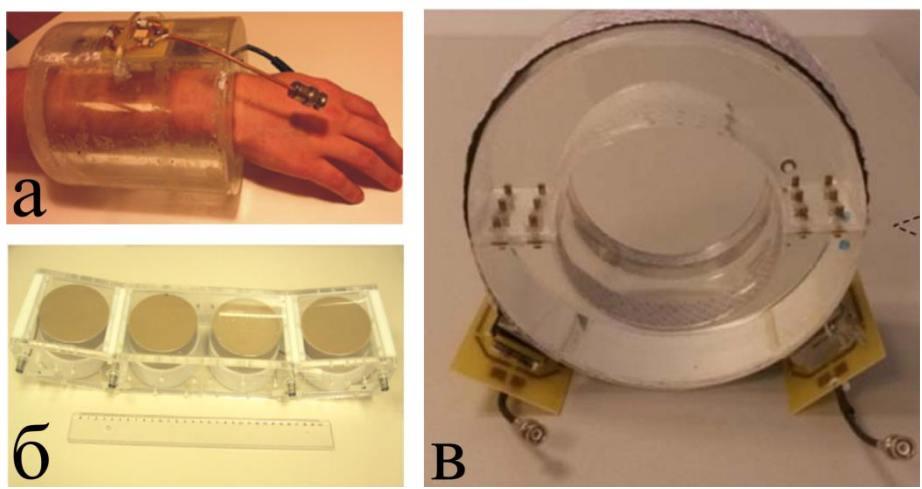

**Рис. 5** (а) резонатор для проведения обследований запястья, (б) массив керамических резонаторов для сверхвысокопольной томографии сердца, (в) резонатор для сканирования коленных суставов на основе разборной емкости с водой.

Резонаторы из керамики обладают более высоким значением относительной диэлектрической проницаемости, чем вода и соответственно, имеют меньшие размеры, необходимые для формирования резонансных структур. Благодаря этому, они могут быть использованы для построения фазового массива (Рис. 5, б), где они выступают в роли антенных элементов [95]. Такая конструкция сочетает в себе достоинства использования диэлектрических резонаторов и стандартный подход формирования многоканального приемного массива антенн. В качестве примера, данная технология позволяет получать в сверхвысоком поле изображение грудной клетки пациента, на котором практически не наблюдается характерного для сверхвысоких полей негативного влияния интерференционных явлений.

В ранее описанном устройстве [94] рамки, возбуждающие резонатор, размещены внутри резонатора и не могут быть отсоединены. В то же время, наличие разъемного соединения внутри или на поверхности катушки облегчает правильное позиционирование объекта, а также позволяет обеспечить относительный комфорт пациенту во время прохождения обследования. Наличие разъема, как правило, изменяет свойства резонатора, что может быть фатально для формирования необходимых мод. Однако нужные моды могут быть восстановлены с помощью включения проводников в структуру резонатора. Так было создано разъемное устройство (Рис. 5, в), в котором проводники позволяют компенсировать неоднородность поля внутри диэлектрика, возникающую в области соединения. Тем самым обеспечивается восстановление резонаторной моды, необходимой для исследования колена [96].

Комбинирование классических методик получения изображений в МРТ с новыми материалами и резонансными структурами открывает возможности по улучшению чувствительности клинических систем, уменьшению времени, необходимому на прохождение обследования, а также снижению радиочастотной нагрузки на пациентов.

## 5. Волноводная МРТ

Как уже было отмечено, в современных клинических высокопольных томографах с величиной индукции статического магнитного поля от 1.5 до 3 Тл типичная РЧ-катушка для возбуждения сигнала ЯМР представляет собой резонатор типа «птичья клетка», впервые предложенный в 1985 году [29]. Конструкция и способ возбуждения мод катушки такого типа уже были описаны ранее. Стоит отметить, что генерируемая такой катушкой круговая поляризация радиочастотного магнитного поля обеспечивает двукратное повышение энергоэффективности МРТ (двукратное снижение потребляемой мощности при той же амплитуде генерируемого РЧ-поля) по сравнению с так называемой седельной катушкой, создающей линейно поляризованное поле $B_1$ и применяемой в ранних МРТ-технологиях. Преимущества катушки типа «птичья клетка» особо заметны в томографах с индукцией статического магнитного поля менее 3 Тл. Однако, при использовании катушек данного типа уже на 3 Тл МР-томографах равномерность возбуждения ЯМР сигнала в теле человека заметно снижается, а потребляемая мощность в два раза больше по сравнению с катушками в томографах с индукцией поля 1.5 Тл. Такое ухудшение характеристик связано с двукратным уменьшением длины волны радиочастотного сигнала и ростом радиочастотных потерь в теле человека в области около 130 МГц – на рабочих частотах 3 Тл МР томографа [97].

Современные сверхвысокопольные доклинические МР-томографы с величиной индукции основного магнитного поля 7 Тл не позволяют единовременно проводить ЯМР-исследование всего тела человека. Несмотря на то, что эти системы обычно оснащены градиентными катушками большого диаметра (около 0,64 м) способными кодировать ЯМР сигнал в области, сравнимой по геометрическим размерам с телом человека, для сверхвысокопольных МР-томографов в настоящее время не разработано универсальной радиочастотной катушки для возбуждения сигнала ЯМР в настолько большой области пространства. Конструкция упомянутой ранее катушки типа «птичья клетка» не может быть адаптирована для томографов с полем 7 Тл. Как уже отмечалось, принцип работы катушки такого типа основан на резонансных условиях течения РЧ тока по контуру двух ее колец и вдоль продольных прутьев, поэтому при фиксированном диаметре и повышении рабочей частоты более чем в два раза резонансная длина прутьев значительно сокращается [97]. Вследствие этого катушка типа «птичья клетка» становится неприменима для возбуждения сигнала ЯМР во всем теле человека на МР-томографах с полем 7 Тл и выше. В настоящее время для получения МР-изображений на сверхвысокопольных томографах используются локальные возбуждающие катушки, специально разработанные для заданной части тела. Такие катушки помещаются непосредственно на поверхность тела пациента и обеспечивают требуемую амплитуду РЧ магнитного поля в ограниченной области. Отсутствие эффективной, универсальной катушки для сканирования больших областей в МР-томографах с индукцией поля 7 Тл и выше является одним из главных препятствий для клинических применений сверхвысокопольных томографов.

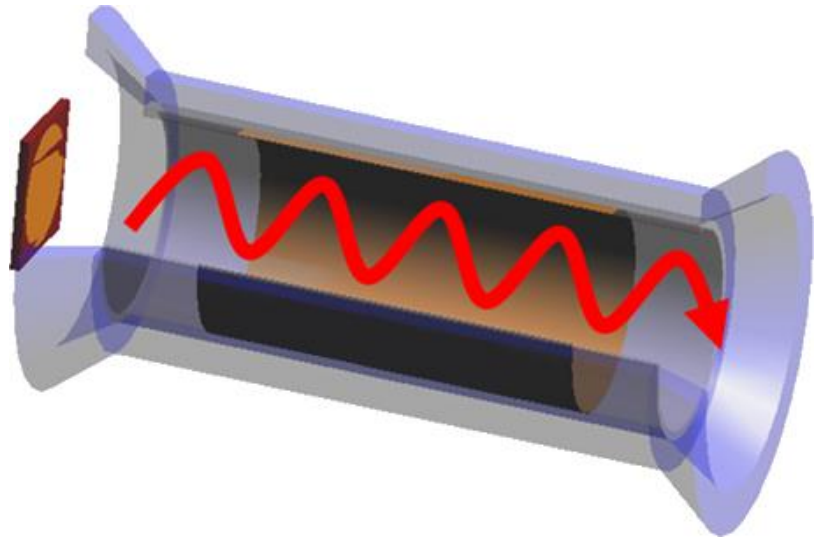

**Рис. 6** Использование туннеля томографа в качестве волновода для проведения исследований человека: радиочастотный сигнал распространяется вдоль оси туннеля, который выполняет функцию цилиндрического металлического волновода на частоте близкой к 300 МГц (рабочая частота 7 Тл МРТ). Этот сигнал возбуждается и принимается антенной, находящейся у левого конца туннеля. Сигнал почти полностью отражается от правого конца, потому что излучение из открытого конца одномодового волновода незначительно.

Относительно малую длину волны радиочастотного сигнала сверхвысокопольных МР-томографов не всегда можно рассматривать, как недостаток. Например, на частоте 300 МГц металлический туннель томографа превращается в отрезок цилиндрического волновода (Рис. 6), который переносит поток радиочастотного электромагнитного поля от антенны, расположенной в одном конце туннеля, к другому концу. Действительно, в соответствии с общеизвестной формулой для частоты отсечки первой фундаментальной моды ($TE_{11}$) полого металлического волновода круглого сечения диметром $d$, заполненного средой с абсолютной диэлектрической проницаемостью $\varepsilon$ и абсолютной магнитной проницаемостью $\mu$

$$f_c = \frac{1.841}{\pi d \sqrt{\mu \varepsilon}},$$

при стандартном диаметре туннеля 0.6 м мы имеем $f_c$ = 293 МГц. На самом деле ввиду частичного заполнения туннеля телом человека, эта частота несколько ниже. Частоты вблизи 300 МГц, соответствующие 7 Тл МРТ, расположены ниже частоты возбуждения следующей ($TM_{01}$) моды, которая равна 381 МГц для пустого тоннеля. Поэтому потери в металле при распространении излучения вдоль туннеля томографа оказываются минимальны именно в области 300 МГц.

Сверхвысокопольную МР-томографию на основе волноводного принципа передачи РЧ-сигнала называют МРТ с использованием бегущей волны или волноводной МРТ [98]. Стоит отметить, что возможность использования волновода в качестве энергетически эффективной системы систем возбуждения и детектирования высокопольного ЯМР была отмечена уже в 1977 году [99]. Однако, размеры предложенных волноводов были несовместимы с использовавшимися на тот момент диаметрами туннелей МРТ, а потому для реализации данного принципа были разработаны специальные линии передачи [99].

Волноводная МРТ обладает несколькими потенциальными преимуществами по сравнению с МРТ, использующей классические радиочастотные катушки и даже антенные решетки с параллельной передачей РЧ сигналов. Во-первых, волноводная технология позволяет получать МР-изображения больших областей, ограниченных только туннелем томографа и градиентными катушками. Во-вторых, удаленное расположение передающей антенны способствует комфорту и безопасности обследуемого человека. С другой стороны, такое размещение антенны обеспечивает пространство для дополнительного оборудования, например, для систем стимуляции, применяемых в функциональной МРТ.

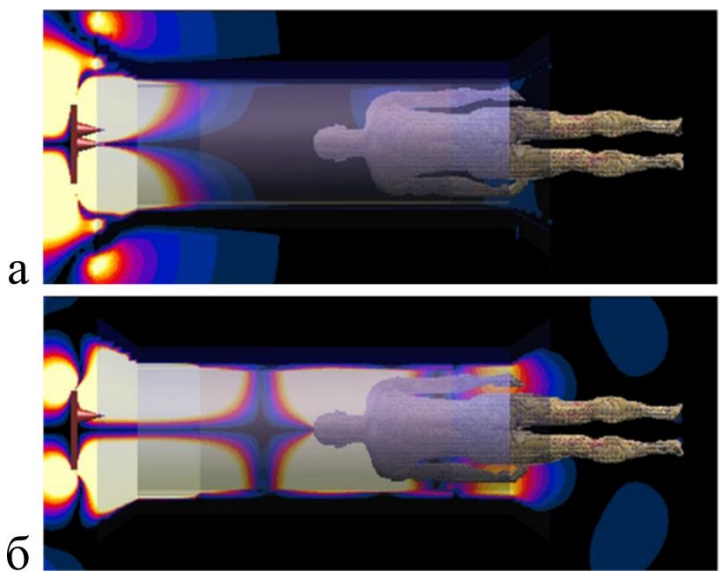

**Рис. 7 Использование МРТ на волноводном принципе для исследований человека возможно только при величине индукции статического магнитного поля порядка 7 Тл и выше, поскольку в этом случае длина волны радиочастотного сигнала сравнима с диаметром туннеля МР томографа. (а) – распределение поля в томографе с индукцией поля 1.5 Тл. Размер туннеля слишком мал в сравнении с рабочей длиной волны томографа (около 4.7 м), поэтому РЧ-сигнал по нему не распространяется. (б) – в томографе с величиной индукции поля 7 Тл РЧ сигнал распространяется вдоль оси томографа, при этом за счет отражения от торцов образуется стоячая волна.**

Использование волноводной технологии возможно только в сверхвысокопольной МРТ с образованием стоячей волны (Рис. 7, б) вследствие внутреннего отражения волны от торцов туннеля. Распространение такой волны в стандартных размеров туннеле томографа на рабочих частотах томографов с индукцией поля меньше 7 Тл невозможно (Рис. 7, а). Наличие стоячей волны необходимо учитывать при расположении пациента в туннеле [100–103]. Однако, от стоячей волны можно избавиться путем согласования торцов туннеля со свободным пространством, для чего надо ввести электродинамические нагрузки по периметру торца. Однако это приведет к потерям энергии РЧ сигнала на поглощение или излучение из торцов. Возможно также управлять положением узла стоячей волны. Для этого электродинамические нагрузки надо вводить внутри туннеля. В качестве таких нагрузок могут выступать продольно ориентированные прямые провода, отстоящие на некоторое расстояние от стенки туннеля (такой нагруженный волновод называют коаксиальным [103]). Необходимо также отметить, что в нагруженном волноводе возможно возбуждение одной-двух мод высшего типа, и их интерференция с фундаментальной модой может быть использована для шиммирования [102,103].

Волноводная МРТ в наше время – это отдельная область научных исследований, целью которых является разработка универсальной катушки для сканирования всего тела человека в сверхвысоких магнитных полях. На настоящий момент уже были продемонстрированы основные потенциальные преимущества метода, основанного на волноводных принципах, позволяющего визуализировать большие объёмы [100,101], а также обеспечивающего возможность манипулировать радиочастотными магнитными полями в теле человека [102,103] и, что весьма немаловажно, совместимого с массивами локальных приемных катушек [104,105]. Однако, отношение создаваемой амплитуды поперечного радиочастотного магнитного поля к потребляемой мощности у антенн, применяющихся в волноводной МРТ, остается небольшим, особенно по сравнению с локальными возбуждающими элементами. Кроме того, у радиочастотных усилителей сверхвысокопольных доклинических МР-томографов пиковые мощности весьма ограничены по сравнению с клиническими МР-томографами. Как следствие, клинические МР-последовательности, большинство из которых требуют высокой амплитуды поперечного радиочастотного магнитного поля, в данный момент несовместимы с волноводной МРТ, что сильно ограничивает её применение при МР-сканировании тела человека в сверхвысоких магнитных полях.

## 6. Функциональная МРТ

Принципиально в понятие функциональной МР-томографии могут быть включены любые исследования, получающие серию МР-изображений того или иного органа в ходе выполнения им своей функции (как в норме, так и при наличии патологии). Однако чаще всего под понятием функциональной МРТ (фМРТ) подразумевается именно МР-исследование работы головного мозга.

Наиболее распространенным методом фМРТ является фМРТ с контрастом по насыщенности крови кислородом (BOLD-контрастом). Данный метод заключается в получении серии изображений головного мозга при выполнении испытуемым заранее сформулированных заданий (и без задания при исследовании функции мозга в состоянии покоя). BOLD фМРТ позволяет обнаружить изменение кровоснабжения тканей головного мозга при переходе из состояния покоя в активное состояние благодаря сопутствующему скачку интенсивности сигнала ЯМР от тканей головного мозга, обусловленному разницей релаксационных параметров тканей при кровоснабжении в состоянии покоя и в активном состоянии.

При переведении фМРТ исследований с томографов, работающих в высоком поле, на томографы, работающие в сверхвысоких полях на чувствительность BOLD метода влияет как изменение амплитуды сигнала ЯМР в сверхвысоком поле, так, очевидно, и смена релаксационных параметром тканей мозга в сверхвысоком поле. Вследствие этого происходит постепенный сдвиг схемы и цели фМРТ сканирования.

Считается, что отношение сигнал/шум при МР-сканировании пропорционально величине индукции магнитного поля (или даже квадратному корню индукции, при учете необходимого увеличения полосы пропускания приёмного тракта при повышении рабочей частоты МР-томографа) и кубу пространственного разрешения [106,107]. В этом приближении переход от стандартного для фМРТ сканирования в поле 3 Тл в поле с индукцией 7 Тл позволяет увеличить изотропное пространственное разрешение в 1.15 или 1.33 раза. С другой стороны, в связи с интенсивной разработкой новых типов РЧ-приёмных устройств для сверхвысоких полей возник вопрос о идеальном отношении сигнал/шум, достижимым для определенного объекта или класса объектов сканирования, в частности для различной степени точности моделей головы человека. Путём численных расчетов было показано и затем подтверждено измерениями (Рис. 8), что создание вокруг выбранного объекта некоторой идеальной конфигурации токов позволяет нелинейно (с показателем степени больше единицы) повышать отношение сигнал/шум при изменении индукции магнитного поля [108–110]. Экспериментальное исследование зависимости отношения сигнал/шум от величины индукции статического поля также показало, что рост нелинеен: отношение сигнал/шум в центре головы человека в поле с величиной индукции 7 Тл превышает соответствующее значение в поле 3 Тл в 3.36 раза, во внешних кортикальных слоях разница менее выражена и превышение отношения сигнал/шум составляет около 2.96 раз [111]. Таким образом, реальное достижимое увеличение разрешающей способности при переходе в сверхвысокое поле составляет до 1.5 раз. В результате этого пространственное разрешение получаемых фМРТ изображений приближается к границе макро- и мезоскопических методов. То есть, если в высоком поле (1-3 Тл) фМРТ регистрирует активацию областей головного с объёмом около 3-5 мм$^3$ [112,113], то при переходе в сверхвысокое поле объём отображаемого вокселя может быть снижен до субмиллиметровых величин [114,115], что позволяет отображать активацию отдельных кортикальных колонок.

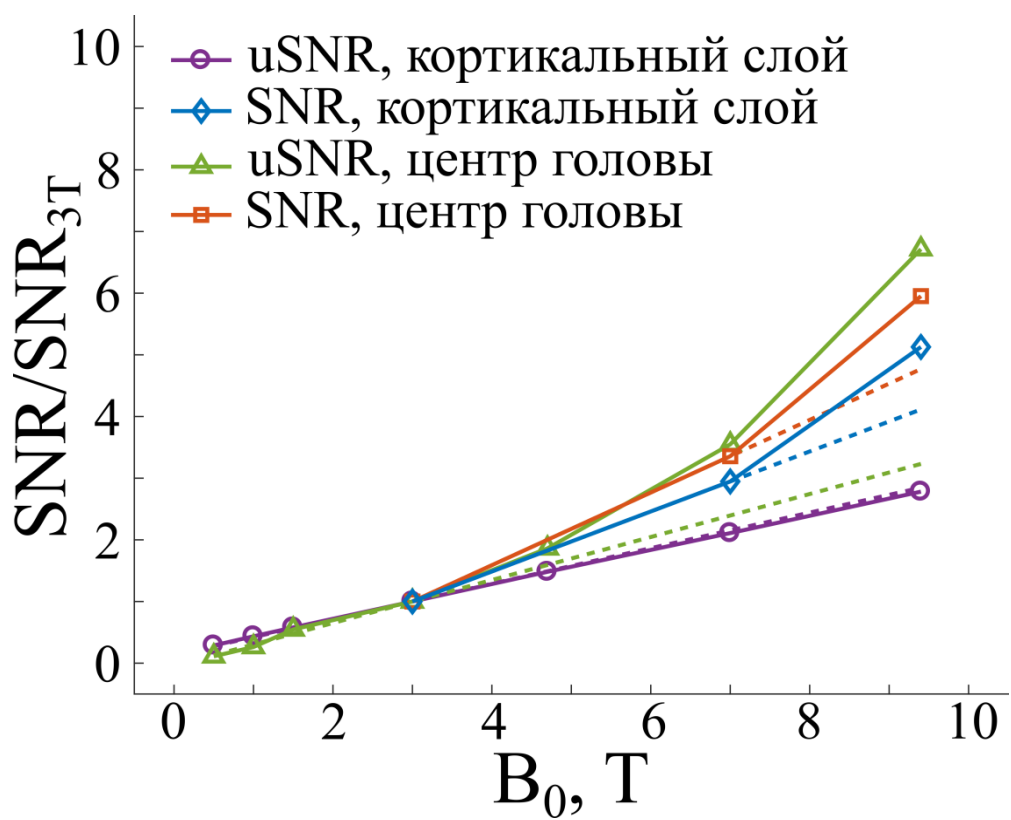

**Рис. 8** Экспериментальные значения отношения сигнал/шум (SNR) и результаты численного моделирования отношения сигнал/шум при создании идеального распределения токов вокруг исследуемого объекта (uSNR), измеренные в кортикальном слое мозга и в центре головы. Пунктирными линиями показана линейная аппроксимация зависимости отношения сигнал/шум в каждом случае. Каждая из зависимостей нормирована на величину отношения сигнал/шум в поле $B_0$=3 Тл. Очевидно, что во всех случаях кроме моделирования uSNR в кортикальном слое рост отношения сигнал/шум при повышении индукции магнитного поля происходит быстрее, чем при линейной аппроксимации.

Ещё более существенным источником фокусировки метода BOLD фМРТ на субмиллиметровых структурах в сверхвысоких полях является относительное изменение релаксационных параметров интраваскулярных и экстраваскулярных кровяных тканей, а также релаксационных параметров ткани мозга, связанных с сосудами различного диаметра. Систематизация данных о релаксационных свойствах крови позволяет построить эмпирические зависимости величины изменения сигнала ЯМР при BOLD фМРТ от диаметра сосуда, насыщенности крови кислородом, а также величины магнитного поля. Кроме того величина BOLD-отклика будет зависеть от используемой импульсной последовательности вследствие различных механизмов релаксации, детектируемых в последовательностях на основе спинового [116] и градиентного [117] эха. Анализ полученных зависимостей показывает, что при использовании последовательности на основе градиентного эха не существует таких условий (Рис. 9, б), при которых возможно превышение сигнала, связанного с малыми сосудами (до 0.02 мм) над сигналом, связанным с крупными сосудами (0.2 мм и выше). В то же время, при использовании последовательности на основе спинового эха (Рис. 9, а) отношение двух этих сигналов достигает максимума при наблюдении в полях от 4 до 8 Тл, причем в полях до 3 Тл отношение не превышает 1 [118]. Таким образом, можно сделать вывод о связи регистрируемого в сверхвысоком поле BOLD-отклика на структурах с характерными размерами в десятки микрон, в то время как в высоком поле наблюдается преимущественно отклик от структур с размерами порядка десятых долей миллиметра.

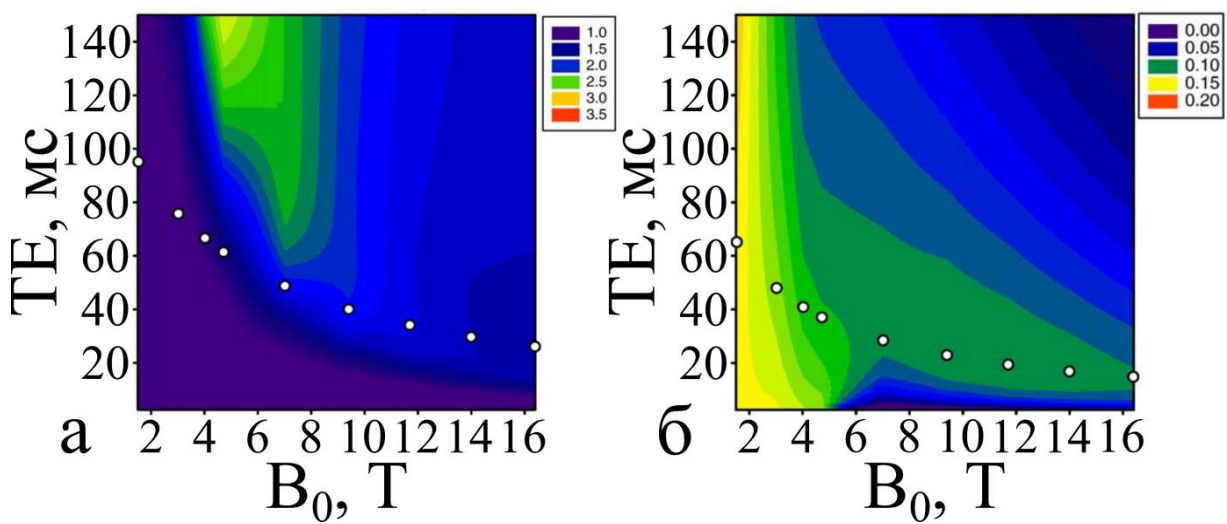

**Рис. 9** Зависимость отношения величины BOLD-отклика, связанного с мелкими сосудами, к величине BOLD-отклика, связанного с крупными сосудами, от величины индукции статического магнитного поля ($B_0$) и от времени наблюдения сигнала эха (TE) для (а) последовательности на основе спинового эха и для (б) последовательности на основе градиентного эха. Стоит отдельно отметить разницу в масштабах двух графиков.

## 7. Диффузно-взвешенная МРТ

Чувствительность к хаотическому молекулярному движению (диффузии) в МРТ является основой для нескольких диагностических и исследовательских методов: диффузно-взвешенной МРТ (DWI), диффузно-тензорного картирования (DTI) [9], картирования спектра диффузии (DSI) [119] и их модификаций [120,121]. В основе методов получения информации о преимущественном направлении диффузии (DTI, DSI и других) лежит получение нескольких карт распределения величины коэффициента самодиффузии (аналогично DWI), отличающихся направлением измерения диффузии а также степенью влияния диффузии на интенсивность регистрируемого сигнала (диффузной взвешенностью). Вследствие этого факторы, влияющие на диффузно-взвешенные изображения, воздействуют также и на данные, получаемые в ходе реализации более сложных методик.

В первую очередь, как и в случае с функциональной МРТ, фактором, влияющим на результаты диффузно-взвешенного МР-сканирования является повышение отношения сигнал/шум с ростом магнитного поля. Это позволяет значительно увеличить пространственное разрешение диффузно-взвешенного сканирования при переходе к сверхвысокопольной МРТ от высокопольной. С другой стороны, отношение сигнал/шум в классических диффузно-взвешенных спин-эховых последовательностях (SE-EPI) зависит от времени поперечной релаксации $T_2$ и в исследуемой ткани, в первую очередь, для белого вещества головного мозга, а также других миелинизированных нервных проводящих путей, представляющих собой основные объекты таких исследований. С ростом магнитного поля наблюдается уменьшение $T_2$, а значит при фиксированном времени регистрации сигнала эха (TE) относительная наблюдаемая величина сигнала падает, что в свою очередь приводит к падению отношения сигнал/шум. В результате учёта релаксационных явлений и других эффектов, влияющих на отношение сигнал/шум, например, увеличения полосы частот приёмного канала (это делается для компенсации артефактов сканирования в сверхвысоком поле и снижения числа шагов так называемого фазового кодирования) отношение сигнал/шум для классической диффузно-взвешенной последовательности в сверхвысоком поле оказывается больше, чем в клинических полях (например, 3 Тл), но только в определенном диапазоне времен TE (Рис. 10). Это преимущество 7 Тл МРТ над 3 Тл МРТ существует при одинаковой полосе пропускания приемного канала, если TE меньше, чем 250 мс. В случае адаптированной полосы пропускания, преимущество сверхвысокольной МРТ исчезает, когда TE превышает 150 мс [122].

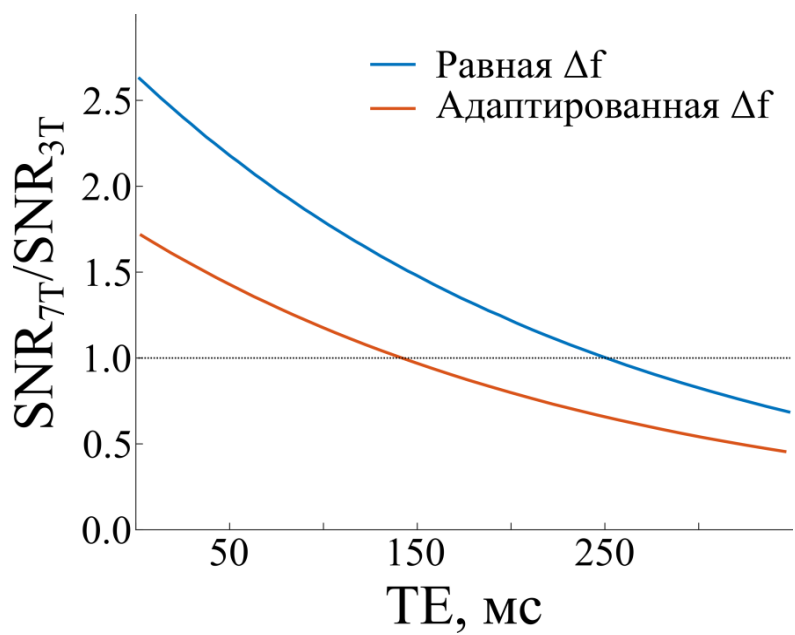

**Рис. 10 Симуляция зависимости отношения сигнал/шум в диффузно-взвешенной импульсной последовательности в поле 7 Тл ($SNR_{7T}$) и в поле 3 Тл ($SNR_{3T}$) в зависимости от времени наблюдения сигнала эха. В симуляциях рассмотрена как ситуация измерения при равной полосе пропускания (Δf) приёмника, так и адаптированной с учетом необходимости увеличения Δf для снижения влияния артефактов на изображение в высоком поле. Пунктиром отмечен уровень равного отношения сигнал/шум в двух симуляциях.**

Нетривиальная зависимость отношения сигнал/шум от величины постоянного магнитного поля осложняется при диффузно-взвешенной томографии различными отношениями сигнал/шум на изображениях с различной диффузной взвешенностью, полученными в рамках одного эксперимента. Использование таких данных для вычисления параметров анизотропной диффузии (фракционной/частичной анизотропии, средней диффузивности и других) приводит к зависимости наблюдаемых значений от величины статического магнитного поля. Так, при повышении поля наблюдается существенное повышение фракционной анизотропии и снижение средней диффузивности [123,124]. Очевидно, что влияние величины магнитного поля на процесс диффузии в МРТ минимально и данный эффект является следствием особенностей процесса измерения диффузии. Действительно, было показано, что сканирование с точным контролем за временем TE и за отношением сигнал/шум приводит к совпадению результатов измерения параметров анизотропной диффузии в высоких и сверхвысоких полях [125,126]. Кроме того, было продемонстрировано [127], что при равных параметрах сканирование в сверхвысоком поле позволяет получить более точные карты распределения параметров анизотропии за счет повышенного отношения сигнал/шум (Рис. 11, а, б).

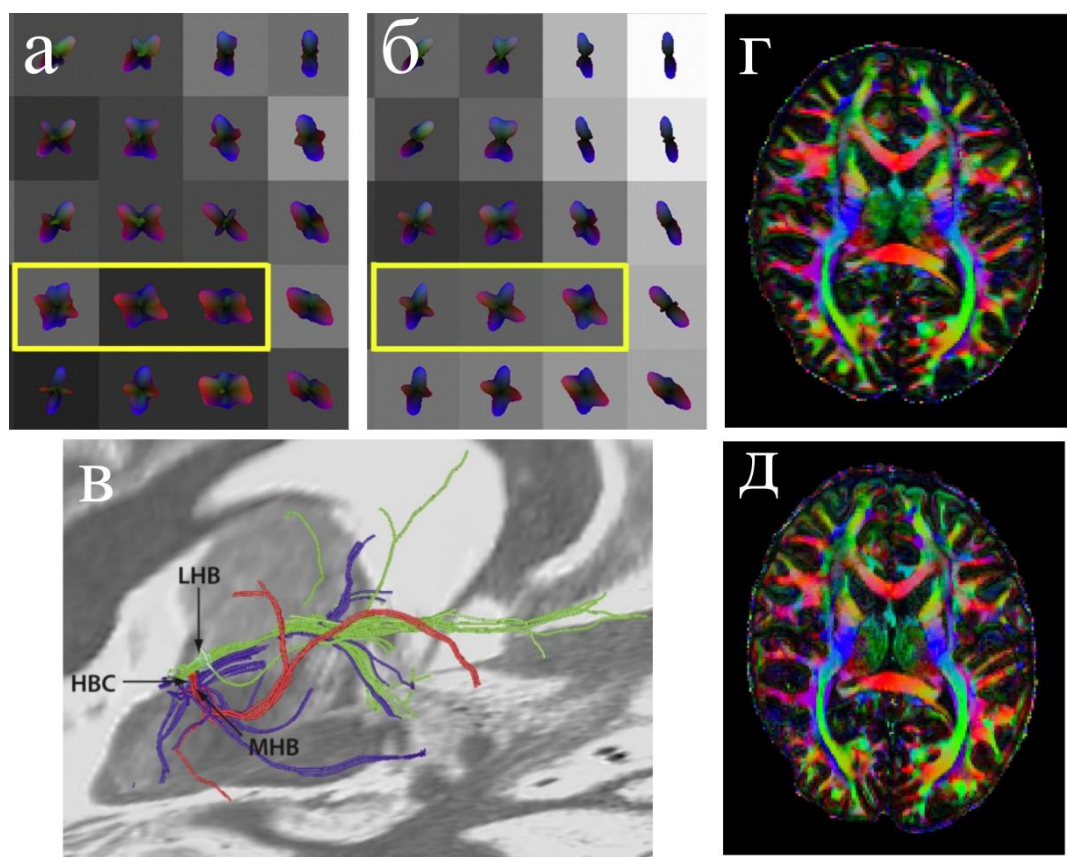

**Рис. 11. Примеры результатов диффузно-взвешенного сканирования в сверхвысоких полях. (а) – картирование угловой компоненты плотности вероятности смещения частиц, характеризующей основные направления диффузии, полученное в поле 3 Тл. (б) – аналогичное картирование, полученное в поле 7 Тл. Распределения, полученные в сверхвысоком поле более отчетливы и позволяют точнее оценить анизотропию диффузии, что наиболее заметно в распределениях в области, отмеченной желтым прямоугольником. (в) – реконструкция нервных волокон, исходящих из латеральной (LHB) и медиальной (MHB) частей поводка эпиталамуса, а также из комиссуры поводка эпиталамуса (HBC) по данным сканирования в сверхвысоком поле. Различие между нервными волокнами удается получить за счет субмиллиметрового разрешения исходных данных. (г) – результаты реконструкции направления нервных волокон во всем мозге (показан один слой) по данным сканирования в высоком поле. (д) – аналогичные результаты, полученные для того же пациента в сверхвысоком поле. Заметна большая детализация изображения и более точная оценка анизотропии диффузии (отмечена яркостью на изображении).**

Повышение отношения сигнал/шум также позволяет использовать для диффузно-взвешенного сканирования в сверхвысоком поле нестандартные последовательности, например, диффузно-взвешенное турбо-спиновое эхо [128], менее подверженное артефактам, связанным с неоднородностью магнитного поля, чем классический SE-EPI метод. Благодаря повышенному отношению сигнал/шум также становится возможным осуществлять прицельное диффузно-взвешенное сканирование миниатюрных областей (таких как, например, поводка эпиталамуса [129]) и исследовать проходящие через них нервные волокна (Рис. 11, в). При сканировании всего головного мозга увеличение отношения сигнал/шум позволяет сократить количество собираемых данных [130] и тем не менее получить значительно более четкие карты анизотропной диффузии (Рис. 11, г, д).

## 8. Перенос намагниченности в ходе химического обмена

При наблюдении ядерного магнитного резонанса в жидкостях растворенное вещество и растворитель могут существовать (в зависимости от температуры, химического состава, частоты наблюдения резонанса и других параметров) в двух режимах: быстрый обмен и медленный обмен. Ситуации быстрого и медленного обмена характеризуется соотношением химического сдвига между резонансами двух популяций ЯМР-активных ядер и скоростью обмена ядрами между ними. В классической ЯМР-спектроскопии ситуации медленного обмена соответствует наблюдение двух спектральных линий от каждой из популяций, а ситуации быстрого обмена – наблюдение одной резонансной линии, усредненной по двум популяциям.

Несмотря на то, что в классической томографии, в основном, под сигналом ЯМР подразумевается спектральная линия свободной воды, наблюдение раствора веществ в режиме медленного обмена может с высокой эффективностью применяться для получения пространственного распределения растворенных веществ. Для этого используется узкополосное насыщение резонансных линий растворенного вещества и регистрация возникающего вследствие обмена уменьшения сигнала свободной воды (метод CEST, chemical exchange saturation transfer).

Эффективность переноса намагниченности (а значит, и чувствительность метода) оценивается различными способами. Самый простой критерий оценки – величина относительного протонного обмена (PTR, proton transfer ratio), вычисляемая, как отношение изменения амплитуды сигнала крупной популяции ядер ($S_w$) к изначальному сигналу от этой популяции ($S_{0w}$). В случае обмена между двумя популяциями с большой разницей равновесных заселенностей (и равновесных констант обмена) PTR можно связать [131] непосредственно с константами релаксации ($R_{1w}$, $R_{1s}$) и обмена ($k_{sw}$) двух популяций (соответствующим индексам $s$ и $w$):

$$PTR = \frac{S_{0w} - S_w}{S_{0w}} = \frac{k_{sw}\alpha x_{CA}}{R_{1w} + k_{sw}x_{CA}}\left[1 - e^{-(R_{1w}+k_{sw}x_{CA})t_{sat}}\right],$$

где $t_{sat}$ – продолжительность импульса, насыщающего резонанс растворенного вещества, $x_{CA}$ – отношение содержания резонирующих ядер, участвующих в обмене, к их общему содержанию в растворителе, и $\alpha$ – величина, зависящая от констант релаксации и обмена и характеризующая эффективность насыщения резонанса ($\alpha = 1$ при полном подавлении пика).

Для характеристики эффективности обмена используется также величина обменного протонного усиления (PTE, proton transfer enhancement), являющаяся PTR, нормированной на $x_{CA}$. Однако и PTR, и PTE не учитывают широкополосного переноса намагниченности и подавления сигнала от растворителя импульсом насыщения. Для учёта этих факторов и зависимости от несущей частоты импульса насыщения вводят величину относительного обмена намагниченностью (MTR, magnetization transfer ratio), которую в свою очередь можно представить в виде составляющих, отвечающих различным источникам изменения наблюдаемого сигнала:

$$MTR(\Delta\omega) = 1 - \frac{S_w(\Delta\omega)}{S_{0w}(\Delta\omega)} = PTR(\Delta\omega) + MTC(\Delta\omega) + DC(\Delta\omega),$$

где MTC – член, отвечающий за широкополосный перенос намагниченности, а DC – за непосредственное насыщение резонанса растворителя импульсом насыщения. Зависимость MTR от разности частот насыщения и резонанса растворителя ($\Delta\omega$) также называют z-спектром или CEST-спектром. Для экспериментального получения величины PTR используют тот факт, что непосредственное насыщение и широкополочный перенос намагниченности симметричны относительно частоты резонанса растворителя, а PTR зависит только от наличия резонансного пика растворителя на частоте облучения, поэтому для большинства веществ будет справедливо

$$MTR_{assym}(\Delta\omega) = MTR(\Delta\omega) - MTR(-\Delta\omega) = \frac{S_w(-\Delta\omega) - S_w(\Delta\omega)}{S_{0w}(\Delta\omega)} = PTR(\Delta\omega),$$

так как пик растворённого вещества отсутствует на частоте $-\Delta\omega$.

Очевидно, что эффективность CEST-метода напрямую зависит от величины $k_{sw}$, так как чем больше ядер, составляющих намагниченность в состоянии насыщения, будет перенесено с растворённого вещества в растворитель, тем сильнее будет изменение сигнала растворителя. Однако необходимость наблюдения z-спектра в условиях медленного обмена (то есть при $\Delta\omega \gg k_{sw}$) требует перехода в большее магнитное поле для наблюдения обмена с высоким $k_{sw}$, так как для двух произвольных веществ $\Delta\omega \sim B_0$. С увеличением $k_{sw}$ падает эффективность насыщения $\alpha$, так как в популяцию ядер растворённого вещества поступает больше ядер из ненасыщенной популяции растворителя. В этом случае повышение $\alpha$ возможно только путём увеличения мощности импульса насыщения, что не всегда возможно *in-vivo* вследствие ограничений на радиочастотный нагрев тканей живого организма. С другой стороны, повышение величины магнитного поля связано с изменением релаксационных характеристик ($R_{1w}$, $R_{1s}$), что в свою очередь значительно повышает наблюдаемый CEST-эффект [132]. Суммарный эффект повышения величины магнитного поля можно рассчитать (Рис. 12) с учётом имеющихся данных о величине релаксационных параметров в различных полях [131].

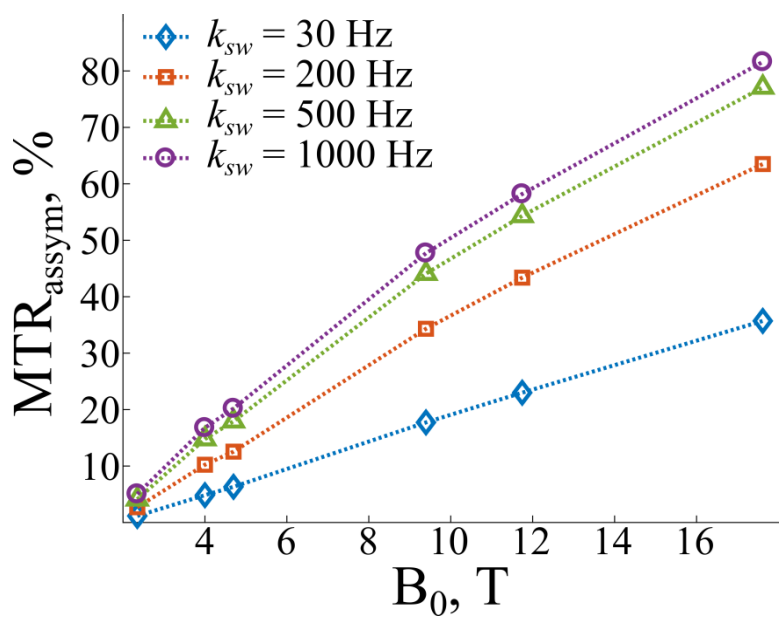

**Рис. 12** Расчет зависимости наблюдаемого CEST-эффекта для одного пика (с химическим сдвигом в 3.5 мд) в зависимости от величины постоянного магнитного поля для корковых тканей мозга мыши. Очевидно, что с повышением магнитного поля становится возможным наблюдение более быстрого обмена и более эффективное наблюдение z-спектра.

Возрастание эффективности CEST-МРТ в сверхвысоком поле позволяет создавать высокоэффективные CEST-агенты на основе гидроксо- и аминогрупп для детектирования раковых образований [133–135], анализа активности центральной нервной системы [136] или повреждений хрящевой ткани [137] часто неприменимые в высоком поле вследствие низких величин химического сдвига [138].

## 9. Восстановление изображений по неполному набору данных

Восстановление исходной функции по неполному набору данных – классическая задача обработки сигналов [139]. Неполноту в данном случае можно понимать, как отклонение от критерия Найквиста, а именно – сбор набора данных со средней частотой меньше удвоенной максимальной частоты сигнала. Усредненность частоты определяется в данном случае, как общее число записанных отсчетов, отнесенное к общей длительности сбора данных. Непосредственное уменьшение частоты оцифровки (с сохранением равномерности снятия отсчетов) однако приводит к ошибкам реконструкции и возникновению артефактов, мешающих клинической интерпретации изображений [140], поэтому для корректной реконструкции используется некогерентная оцифровка и нелинейные процедуры реконструкции, включаемые в группу алгоритмов опознания по сжатию (compressed sensing) или $l_1$-оптимизации, так как с математической точки зрения используемые алгоритмы служат для минимизации $l_1$-нормы разреженного представления изображения, то есть поиска такого $x$, что достигается

$$min\|\Psi x\|_1$$

при условии

$$\|FT(x) - y\|_2 < \varepsilon,$$

где *x* – искомое реконструированное изображение, Ψ – преобразование, переводящее *x* в разреженный вид, *FT* – преобразование Фурье (или при необходимости иное преобразование, используемое для реконструкции изображения), *y* – оцифрованный сигнал, а ε – параметр, характеризующий точность реконструкции [140].

Опознание по сжатию используется в МРТ благодаря избыточной информации, содержащейся в фазовом пространстве. Исключение этой информации из выборки позволяет сократить количество шагов оцифровки и значительно сократить время МР-исследования без видимой потери качества изображений. Применение алгоритмов опознания по сжатию к томографии в высоких полях (до 3 Тл) позволяет реализовать исследования, требующие значительного разрешения, высокого отношения сигнал/шум и при этом кратчайшего времени сбора данных, например отдельные виды динамического контрастного усиления [141–143], ангиографии [144–146] и другие приложения [147–149]. Требование высокого отношения сигнал/шум при высоком разрешении часто автоматически выполняется в сверхвысокопольной (более 3 Тл) МРТ благодаря повышенному сигналу и использованию более мощных градиентных импульсов, однако при этом повышению пространственного разрешения сопутствует длительное время сбора данных, что существенно снижает временное разрешение эксперимента. Таким образом, переход в сверхвысокое поле становится менее эффективным без использования методов повышения временного разрешения, которые часто основаны на алгоритмах опознания по сжатию. В частности, была показана эффективность использования в сверхвысоком поле алгоритмов опознания по сжатию применительно к ангиографии сверхвысокого разрешения [150–153], сверхбыстрым фМРТ исследованиям [154], к визуализации течения жидкости на микроскопическом уровне [155], а также гетероядерным исследованиям [156].

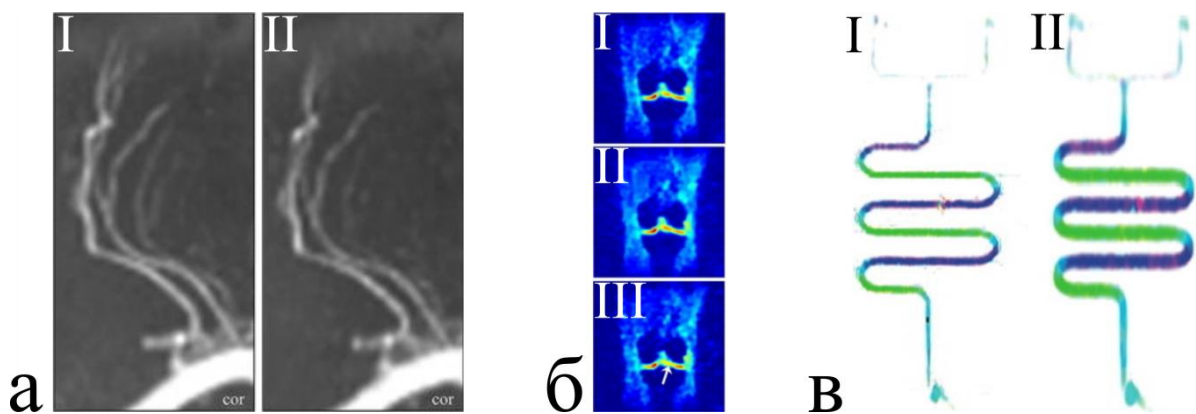

**Рис. 13** Результаты применения алгоритмов опознания по сжатия в сверхвысоком поле. а – время-пролетная ангиография со сверхвысоким разрешением, I – изображение, полученное путем классического сбора данных (время получения ≈ 9.5 минут), II – с использованием опознания по сжатию (время получения ≈ 2.5 минуты). б – карты распределения натрия в коленном суставе. I – полученные с классическим радиальным сбором данных (время получения ≈ 17 минут), II – с некогерентным сокращением количества полученных данных вдвое, III – втрое (стрелка показывает область, в которой наблюдалось отклонение концентрации от контрольного эксперимента). в – карта скоростей в микроканалах, I – полученная с использованием опознания по сжатию (разрешение 256×256 пикселей), II – полученная классическим методом за то же время (разрешение 63×63 пикселя)

Для время-пролетной ангиографии сосудов головного мозга человека со сверхвысоким разрешением (при размере вокселя 0.28×0.28×0.36 мм$^3$) некогерентный сбор данных и реконструкция с использованием разности соседних элементов в качестве Ψ позволяет получить почти четырехкратный выигрыш во времени (Рис. 13, а). При получении карт распределения натрия в суставах использование опознания по сжатию (при опущении от половины до трех четвертей собранных данных и без использования разреживающего преобразования, т.е. $\Psi=1$) наблюдалась успешная реконструкция при сокращении времени сбора данных вдвое, однако при более высоких степенях сжатия были заметны отклонения от вычисленной концентрации натрия в отдельных областях изображений (Рис. 13, б). При визуализации карты скоростей жидкости на чипе (Рис. 13, в) удалось достичь сокращения времени сканирования в 16 раз при сохранении достоверной карты скоростей жидкости в микроканалах. В данном случае в качестве разреживающего преобразования $\Psi$ использовалось разложение в базис по вейвлет-функциям (конкретнее, по функциям-симмлетам). Таким образом, в зависимости от выбранного объекта исследования и алгоритма сбора данных и реконструкции опознание по сжатию позволяет либо сократить время эксперимента до 16 раз без потери качества, либо провести эксперимент с повышенным разрешением за тот же промежуток времени.

## 10. Распознавание по МР-отпечатку

Переход от классической стратегии сбора данных к более эффективным и современным методам происходит не только благодаря использованию псевдохаотического заполнения k-пространства (см. главу Восстановление изображений по неполному набору данных). Сам процесс проведения МР-сканирования может быть подвержен существенной модификации.

В настоящее время основной массив данных МР-томографии собирается в рамках протоколов сканирования, состоящих из МР-последовательностей. Каждая из последовательностей согласно её типу и параметрам обычно даёт одну или несколько серий изображений с определённым соотношением интенсивностей сигнала ЯМР от различных тканей – так называемую «взвешенность изображения». Интенсивность сигнала на подобных взвешенных изображениях определяется в первую очередь релаксационными характеристиками тканей, однако сами релаксационные параметры тканей при этом не измеряются. При необходимости провести такое исследование используют относительно длительные последовательности, изолирующие влияние различных параметров на интенсивность ЯМР сигнала, а затем из набора подобных изображений производится вычисление таких параметров, как время продольной или поперечной релаксации, или коэффициент самодиффузии [157–161]. С другой стороны, требование изоляции влияния не измеряемых в конкретном эксперименте параметров означает, что при произвольном подборе параметров сканирования вклад в интенсивность сигнала ЯМР будут вносить многие из характерных параметров тканей. Этот факт приводит к смене парадигмы численной оценки релаксационных (таких как времена релаксации $T_1$, $T_2$ или $T_2^*$) и других характеристик (например, коэффициента самодиффузии, магнитной восприимчивости и прочих) тканей в МРТ и появлению методики распознавания МР-отпечатков (MR-fingerprinting) биологических тканей [162].

Метод распознавания МР-отпечатков основывается на зависимости сигнала ЯМР, регистрируемого в ходе произвольной импульсной последовательности от полного набора релаксационных и иных характеристик отображаемой ткани. Так как сигнал ЯМР зависит именно от всего набора свойств ткани, то для определённой последовательности импульсов величина сигнала, регистрируемого в некоторый выбранный момент времени, может быть характерна для ткани с присущим только ей набором свойств. Очевидно, что при заданном соотношении параметров такни и параметров последовательности единичный сигнал может быть идентичным для двух типов тканей с различным набором свойств, поэтому для точного сопоставления параметров сигнала и характеристик ткани используют, во-первых, последовательность с псевдохаотическим набором импульсов, а во-вторых большой набор последовательных сигналов ЯМР. Такой набор (даже при совпадении величин сигнала от двух тканей в одной временной точке) будет уникальным для каждого набора измеримых методом ЯМР параметров ткани, а значит, может служить для её идентификации, также как отпечаток пальца может служить для идентификации личности человека (отсюда и название методики – распознавание по МР-отпечаткам).

Существенной задачей распознавания МР-отпечатков является сам процесс идентификации серии сигналов. Для идентификации отпечатка используются массивы симулированных серий сигналов (словари), в которых каждая серия соответствует определенному набору измеряемых характеристик тканей. Совпадение измеренного сигнала со словарным позволяет идентифицировать ткань и получить численные значения её релаксационных и иных параметров. Очевидно, что количество и разброс параметров, а также разрешение словаря в параметрическом пространстве определяют как скорость поиска совпадения, так и количество компьютерной памяти, требуемое для работы со словарем. Для сокращения времени поиска и уменьшения требований применяются различные алгоритмы сжатия и поиска совпадения [162–168].

Распространенными переменными для составления словарей являются времена релаксации, относительная спиновая плотность и величина статического поля $B_0$ (или степень его неоднородности для учёта нерезонансного поведения намагниченности). При сканировании в высоких полях такой набор параметров позволяет эффективно определить величины времен релаксации тканей головного мозга в норме и при патологии [169], тканей брюшной полости [170] миокарда [171] в норме. Включение в протокол сканирования измерения коэффициента самодиффузии позволило обнаруживать и оценивать стадии рака предстательной железы [172].

Неоднородность радиочастотного магнитного поля, возникающая при переходе в сверхвысокое поле, негативно влияет на точность распознавания тканей по МР-отпечаткам. Поэтому при переходе в поля с величиной индукции 7 Тесла и более проблема неоднородности усиливается и требуется фундаментальный подход к её решению. В рамках распознавания МР-отпечатков таким подходом является включение в словарь симулированных сигналов дополнительных измерений, соответствующих параметрам, отвечающим за возможные вариации радиочастотного поля используемых в МР-томографе РЧ-катушек.

Непосредственная реализация коррекции неоднородности РЧ-поля состоит во включении в словарь отпечатков отклонения от требуемого угла поворота, как одного из параметров (и соответственно, как одного из дополнительных измерений словаря) [170]. Измерения времен релаксации фантомов с подобной коррекцией и без неё показали, что наиболее чувствительными к неоднородностям РЧ поля являются измеренные времена $T_2$, в то время как точность распознавания времен $T_1$ практически не изменяется при отсутствии коррекции неоднородностей переменного поля (Рис. 14). Ещё большей точности измерения можно добиться путём введения в последовательность, используемую для генерации МР-отпечатка, элементов повышающих чувствительность сигнала к величине переменного магнитного поля. Таким элементом, например, может служить серия импульсов с резкой сменой угла отклонения, нарушающая течение классической квазистационарной последовательности для регистрации МР-отпечатка [173].

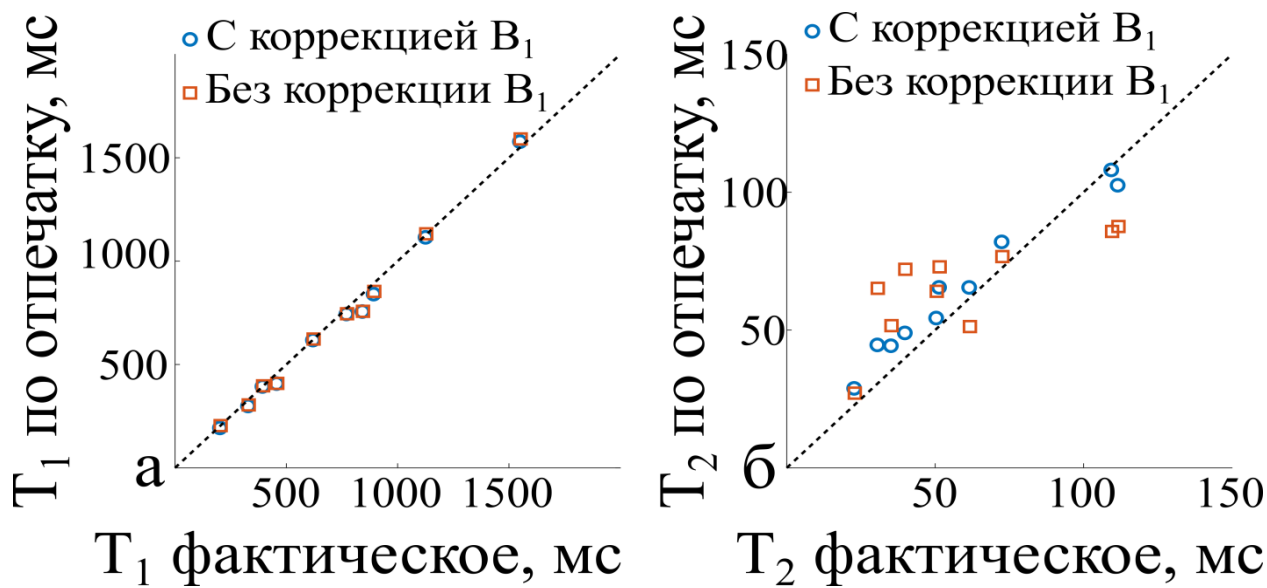

**Рис. 14 Результаты измерения времен релаксации методом распознавания МР-отпечатков. (а) – измерение времени $T_1$, (б) – времени $T_2$. Фактическое время релаксации определено классическими ЯМР-методами измерения времен релаксации. Точность измерения методом распознавания отпечатка определяется близостью точки на графике к диагонали (отмечена пунктиром). Можно заметить, что коррекция неоднородностей РЧ-поля не влияет на точность измерения продольной релаксации, однако позволяет получить лучшее совпадение с классическими методами при измерении поперечной релаксации.**

Вне зависимости от метода, однако, проблема коррекции неоднородности РЧ-поля путём включения её в качестве параметра в словарь распознавания заключается в невозможности скорректировать высокую неоднородность поля. Значительное отклонение фактического переменного магнитного поля от желаемого может приводить наличию областей, в которых радиочастотные импульсы МР-последовательности для создания МР-отпечатков неспособны отклонить намагниченность от равновесного положения на существенный угол (более 5 градусов). В этом случае отпечатки веществ с разными временами релаксации становятся неразличимы. Для устранения этой неоднозначности можно использовать свойства распределения магнитного РЧ-поля различных мод катушки типа «птичья клетка», используемой для генерации магнитного РЧ-поля в большинстве томографов. Например, минимумы поля (представляющие проблемные для распознавания области) вырожденных мод $CP_1$ и $CP_2$ катушки [174] значительно разнесены в пространстве. Вследствие этого возбуждение различных мод катушки в ходе одной последовательности позволяет сгенерировать уникальный отпечаток вне зависимости от наличия минимума поля в точке нахождения объекта, так как расположение минимума одной моды с положением объекта будет компенсировано максимумом другой моды. Дальнейшее включение экспериментально измеренных распределений поля на двух модах катушки в словарь МР-отпечатков позволяет избежать ошибок распознавания, связанных с высокой неоднородностью РЧ-поля [175]. Кроме того данная методика позволяет проводить МР-томографию в анатомических областях, где неоднородность РЧ-поля вызвана не свойствами мод возбуждающей катушки, а наличием металлических объектов, например, хирургических имплантатов (Рис. 15).

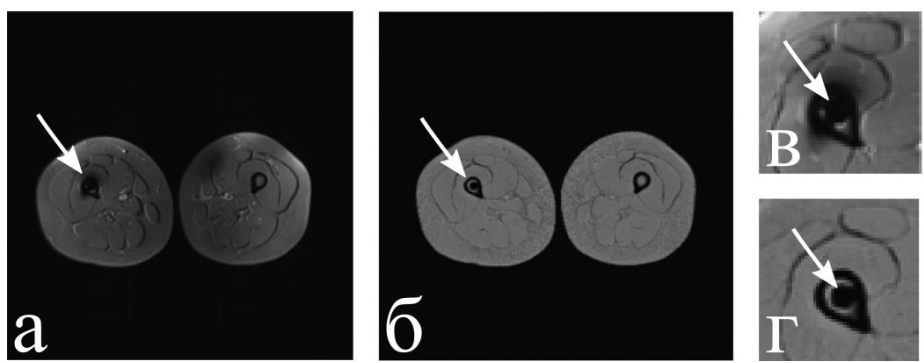

**Рис. 15 МР-изображения конечностей человека с хирургическим имплантатом, полученные с помощью (а, в) классической последовательности инверсии-восстановления со сбором данных методом турбо-спинового эха и (б, г) методом снятия МР-отпечатка с переменным возбуждением различными модами передающей катушки. На общих изображениях (а, б) видно, что артефакты от наличия имплантата присутствуют не только в области его нахождения (указано стрелкой), но и в контралатеральной конечности. Увеличенные изображения места нахождения имплантата (в, г) показывают, что методом снятия МР-отпечатка можно получить изображение без артефактов даже в непосредственной близости от имплантата.**

## 11. Гетероядерная МРТ

Магнитно-резонансная спектроскопия (МРС) – это неионизирующий метод, позволяющий детектировать химический состав и наблюдать метаболизм тканей живого организма в естественных условиях функционирования. Метод *in-vivo* МРС является эффективным инструментом для неинвазивного обнаружения метаболитов в клетках, изучения скорости метаболизма, энергетического обмена, нейротрансмиссии в мозге или других органах, а также изменений этих процессов, вызванных физиопатологическим вмешательством.

МРС на ядрах водорода является самым распространенным видом МРС-исследований, так как все необходимое для нее оборудование уже имеется на любом магнитно-резонансном томографе. Однако помимо водорода в живых организмах присутствует еще несколько видов ядер, дающих магнитно-резонансный отклик (Таблица 1). Детектирование сигнала ЯМР на этих ядрах – задача гетероядерной МРС. Особое внимание в гетероядерной *in-vivo* МРС уделяется следующим ядрам: фосфору ($^{31}$P), углероду ($^{13}$C), кислороду ($^{17}$O), натрию ($^{23}$Na) и фтору ($^{19}$F).

**Таблица 1 Изотопы, использующиеся в МР-спектроскопии живых организмов, их гиромагнитное отношение и распространенность в нормальных условиях.**

| Ядро | Гиромагнитное отношение, МГц/Тесла | Распространенность изотопа, % |
|---|---|---|
| $^{1}$H | 42.58 | 99.99 |
| $^{13}$C | 10.71 | 1.11 |
| $^{17}$O | -5.77 | 0.04 |
| $^{19}$F | 40.05 | 100 |
| $^{23}$Na | 11.26 | 100 |
| $^{31}$P | 17.24 | 100 |

Фосфор в организме является основной составляющей высокоэнергетических соединений таких как АТФ, синтезируемый внутри митохондрий, и фосфокреатин. Вследствие этого с помощью фосфорной МРС возможно изучение биоэнергетических процессов в мозге, сердце, мышцах и иных органах [176–178]. Кроме того, фосфорная МРС позволяет определять и другую важную физиологическую информацию, в частности концентрацию свободного магния, внутриклеточный pH, а также обнаружить промежуточные продукты метаболизма фосфолипидов. МРС на ядрах углерода является единственным методом, который позволяет неинвазивно наблюдать нейроэнергетические процессы и циркуляцию нейромедиаторов путем локализации ключевых ферментов и метаболитов в нейронах и нейроглиях [179]. Также в настоящее время демонстрируются успешные результаты применения МРС на ядрах кислорода к исследованию кислородного обмена в клетках и перфузии в аэробных органах, а также к оценке скорости кислородного обмена и его изменений [180]. Регистрация ЯМР-отклика от ядер натрия методами МРС потенциально может быть полезна во многих областях, включая диагностику инсульта, характеризацию опухолевых образований (Рис. 16, а), раннее обнаружение остеоартрита и оценку целостности хряща в суставе, а также оценку функции мышц и почек [181,182]. Наконец, фтор, концентрация которого в живых организмах близка к нулю, входит в состав нетоксичных и химически инертных соединений, используемых в широком диапазоне биомедицинских применений, включая анестетики, химиотерапевтические агенты и молекулы, служащие для эффективного растворения дыхательного кислорода, использующиеся в заменителях крови [183,184]. Вследствие этого МРС на изотопе $^{19}$F является эффективным и неинвазивным методом визуализации распространения этих соединений в организме (Рис. 16, б).

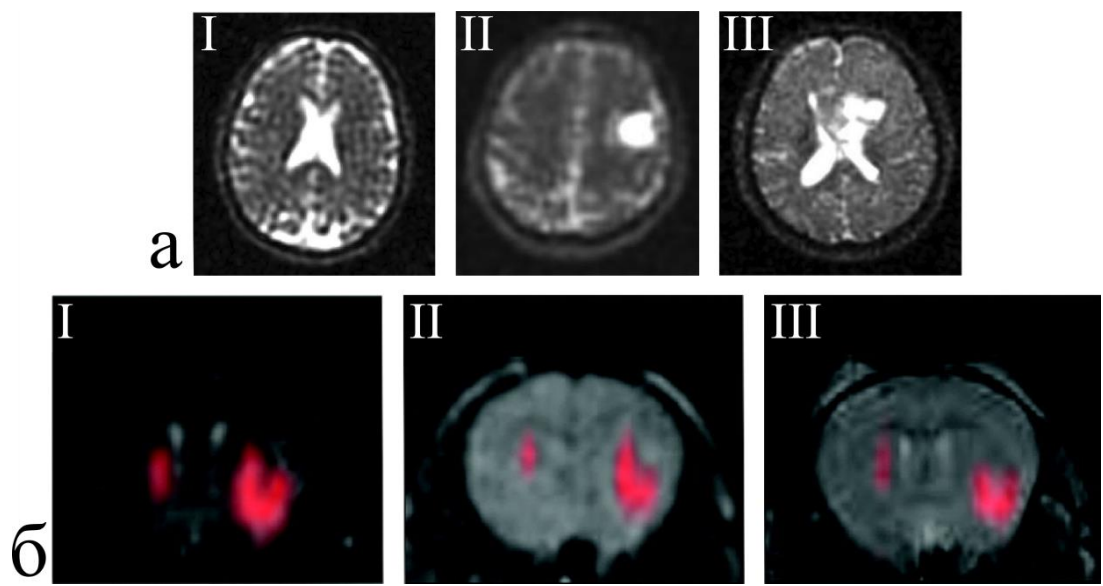

**Рис. 16 Гетероядерные изображения, полученные в сверхвысоких полях. (а) – $^{23}$Na изображения мозга человека. полученные в поле 7 Т. (I) головной мозг человека в норме, (II) головной мозг с ганглиоглиомой в левой лобной доле, (III) головной мозг с глиобластомой левой лобной доли. (б) – карты распределения помеченных изотопом $^{19}$F нервных стволовых клеток (розовым на фоне классического $^{1}$H изображения в шкале серого), имплантированных мыши, полученные в поле 9.4 Т через (I) 1 час, (II) 3 дня и (III) 1 неделю после операции. Карты позволяют оценить жизнеспособность имплантированных клеток с течением времени.**

Основным недостатком гетероядерной МРС является ее относительно низкая чувствительность по сравнению с МРС на ядрах водорода. Это обусловлено совокупностью следующих факторов: низкой концентрацией исследуемых соединений, низкой распространенностью некоторых изотопов, а также пониженным гиромагнитным отношением исследуемых ядер по сравнению с ядрами водорода (Таблица 1). Поэтому переход к сверхвысоким магнитным полям является естественным способом повысить чувствительность гетероядерной МРС. Однако, для разных ядер повышение основного магнитного поля имеет свои последствия в зависимости от типов релаксационных механизмов характерных для этих ядер [185,186]. Тем не менее в целом эффект повышения чувствительности гетероядерной МРС при переходе к сверхвысоким полям сохраняется.

В отличие от МРС на ядрах водорода (источника сигнала обычной магнитно-резонансной томографии), проведение гетероядерной МРС возможно только в случае наличия отдельной радиочастотной цепи возбуждения/приема сигнала и специальных катушек, так как рабочая частота гетероядерных МРС экспериментов определяется исследуемым ядром. Кроме того, в $^{13}$C и $^{31}$P МРС необходимо использовать протонную развязку (proton decoupling) для повышения качества и разрешения получаемых спектров, так как гетероядерные взаимодействия с ядрами водорода через электронные облака усложняют интерпретацию спектров и уширяют резонансные линии. Протонная развязка достигается путем радиочастотного облучения исследуемой ткани на резонансной частоте ядер водорода во время детектирования сигнала от гетероядер, что приводит к увеличенной РЧ нагрузке на объект исследования, в особенности на сверхвысоких полях [187]. Также для $^{13}$C и $^{31}$P МРС зачастую приходится использовать относительно сложные методики переноса намагниченности, что позволяет изучать соединения, содержащие эти ядра, однако детектируя при этом сигнал от ядер водорода. Благодаря этому повышается чувствительность и специфичность эксперимента [188,189]. Для редких изотопов ($^{13}$C) активно исследуются методы гиперполяризации [190], позволяющие повысить чувствительность на несколько порядков. Из-за множества технических сложностей, связанных с созданием и сохранением неравновесной поляризации только быстрые и динамические метаболические пути пока могут быть изучены с помощью МРС гиперполяризованных ядер углерода.

Регистрация сигнала ЯМР от гетероядер не позволяет получать анатомической информации высокого разрешения в силу очень низкой концентрации или полного отсутствия в организме исследуемых изотопов, поэтому гетероядерные исследования всегда необходимо совмещать с классической МР-томографией для правильной локализации гетероядерного сигнала. Кроме того требуется наличие калибрующего сигнала для настройки эксперимента и количественного анализа его результатов [191]. Гетероядерные спектры имеют очень широкий диапазон химических сдвигов резонансных линий (например, для фосфора это 30 мд, а для углерода более 200 мд), что приводит к значительным ошибкам при пространственной локализации источника гетероядерного МР-сигнала. Одной из особенностей гетероядерной МРС (наряду с техническим усложнением методик МРС, ведущим к повышению информативности получаемых данных) в сверхвысоких магнитных полях становится возможность адаптации РЧ устройств изначально разработанных для МРС на ядрах водорода, но в более низких полях, что продиктовано совпадением резонансных частот редких ядер в сверхвысоких полях и резонансной частоты $^1$H в низком поле.

Таким образом, гетероядерная МРС является важным дополнением к классической магнитно-резонансной томографии, так как позволяет неинвазивно получать на молекулярном и клеточном уровнях уникальную информацию, связанную с физиологией и патологией живых тканей. Гетероядерная МРС особо актуальна при проведении исследований в сверхвысоких полях в связи с повышенной чувствительностью и улучшенным спектральным разрешением, а также в связи отсутствием проблем, связанных с созданием эффективных радиочастотных устройств для регистрации сигнала от редких ядер. Однако в настоящий момент из-за ограниченного числа сверхвысокопольных томографов применение гетероядерной МРС в теле человека ограничено исключительно научными исследованиями.

## 12. Заключение

Рассмотрение особенностей МРТ в сверхвысоком поле показывает, что новые задачи, характерные для технологий со статическими магнитными полями выше 3 Тл, порождают значительное количество инженерных новых решений, обладающих фундаментальной научной и технической новизной. Так, проблема неоднородности радиочастотного магнитного поля, возникающая из-за явлений интерференции первичного и вторичного РЧ-излучения решается либо созданием новых приемо-передающих РЧ-катушек, работающих в режиме многоканального сканирования (причем может быть реализован как статический, так и динамический подход к управлению каналами), либо с использованием метаматериалов, либо с использованием диэлектрических подкладок, либо переходом к катушкам на основе диэлектрических резонаторов, либо, наконец, принципиально новым путём – волноводной МРТ. Сложность и взаимосвязь новых задач приводит к появлению комплексных решений, позволяющих преодолеть сразу несколько затруднений на пути к доступной сверхвысокопольной клинической томографии. Так, использование метаматериалов и материалов с высокой диэлектрической проницаемостью открывает возможности как к снижению РЧ-нагрузки на пациента, так и к повышению однородности поля передачи, чему также способствуют использование распознавания по МР-отпечатку и метод распознания по сжатию. Последние позволяют значительно сократить время исследования вследствие неполного сбора данных.

С другой стороны, новые возможности сверхвысокопольной томографии ведут к существенному расширению спектра применения МРТ. Повышение пространственного и функционального разрешения фМРТ позволяет осуществлять наблюдение структур мозга ранее доступных только инвазивным методам. Повышение эффективности переноса намагниченности ведет к созданию более эффективные методы диагностики заболеваний. Использование гетероядерного МРТ также увеличивает количество решаемых при помощи МРТ диагностических задач, а кроме того позволяет исследовать процессы метаболизма непосредственно в живом организме. Распознание по отпечатку способствует переходу от качественной оценки МР-изображений к количественной, что в свою очередь приводит к повышению эффективности клинической диагностики.

Данный обзор сверхвысокопольной МРТ, убедительно показывает, что, несмотря на то, что этот метод диагностики до сих пор не вошёл в клиническую практику, в мире интенсивно ведутся работы по преодолению существующих технических сложностей и идёт поиск новых диагностических возможностей, которые сверхвысокие магнитные поля открывают для современной медицины.

## Благодарность

## Список литературы